%% file: main.tex
\documentclass{article}

\usepackage[english]{babel}
\usepackage{longtable}

\usepackage{placeins}

\usepackage[table,dvipsnames]{xcolor}
\usepackage{booktabs}    
\usepackage{subcaption}  
\usepackage{tikz}
\usetikzlibrary{arrows.meta, positioning, calc}
\usepackage{multirow}
\usepackage{authblk}
\usepackage{algorithm}
\usepackage{algpseudocode}

\usepackage[letterpaper,top=2cm,bottom=2cm,left=3cm,right=3cm,marginparwidth=1.75cm]{geometry}

\usepackage{amsmath}
\usepackage{graphicx}
\usepackage[colorlinks=true, allcolors=blue]{hyperref}
\usepackage{bbm}

\title{Robust Bayesian Sequential Borrowing for Multi-Population Clinical Programmes}

\author[1,3]{Erik Hermansson}
\author[2]{Lynn Dunsire}
\author[1]{David Svensson}
\author[3]{Thomas Jaki}

\affil[1]{R\&I Biometrics and Statistical Innovation, BioPharmaceuticals R\&D,AstraZeneca, Gothenburg, Sweden}
\affil[2]{R\&I Biometrics and Statistical Innovation , BioPharmaceuticals R\&D, AstraZeneca, Cambridge, UK}
\affil[3]{Faculty of Informatics and Data Science, University of Regensburg, Regensburg, Germany}
\begin{document}
\maketitle

\begin{abstract}

We introduce Robust Bayesian Sequential Borrowing (RBSB), a framework for extrapolating evidence across adjacent subgroups in multi-population clinical programmes where studies are conducted in sequence and populations are ordered by clinical proximity. Conventional approaches weight all historical sources uniformly or exclude distant populations entirely, failing to reflect the natural gradient of similarity in such programmes. RBSB encodes the programme order through path-dependent borrowing via robust mixture priors that combine an informative component with a unit-information component to guard against prior-data conflict. Posterior weights, derived in closed form from marginal likelihood ratios, provide transparent dynamic attenuation when heterogeneity arises between sequential populations. The framework supports prospective evaluation of Bayesian Type~I error, power, and extends naturally to assurance at both the study and programme level. Simulation studies demonstrate superior false-positive control relative to full pooling, while preserving substantial efficiency gains over standalone analyses. A case study of the START trial illustrates the approach across adult, adolescent, and paediatric populations. RBSB offers a practical, regulator-aligned method for disciplined evidence borrowing that exploits temporal and biological proximity while preventing implausible extrapolation across distant populations.

\end{abstract}

\section{Introduction}

Drug development across age groups presents scientific and statistical challenges. Practical and ethical constraints often limit the size of paediatric trials, resulting in smaller and potentially underpowered studies relative to adult development programmes. Evidence extrapolation has often relied on models that pool findings from more robust adult studies to support conclusions in younger groups. However, such approaches can miss the fact that in programmes with multiple age groups, some groups are only partly comparable and the degree of similarity varies between groups. Consequently, there is a need for a statistical approach that allows appropriate borrowing of information from more exchangeable populations, applied in a structured and sequential manner to adjacent populations, whilst reducing the influence of data from populations with lower exchangeability.


Recent work has examined borrowing and planning in sequential adult-paediatric programmes, including Campbell et al. \cite{Campbell2024}, who contrast alternative trial design strategies for whole-of-life inclusion and staged development. Complementary methodological advances include the EXNEX framework of Neuenschwander et al. \cite{EXNEX}, which enables stratum-specific borrowing by combining exchangeable and non-exchangeable components, in contrast to standard hierarchical models that assume full exchangeability across strata. In parallel, Emerson et al. \cite{bayes_gsd} evaluate Bayesian group sequential designs that accommodate prior information while allowing interim decision-making within a coherent Bayesian framework.

In 2026, the FDA released a draft guidance on the use of Bayesian methods in clinical trials. The guidance notes that when there is a difference in exchangeability between populations, not all data sources are equally relevant, accordingly, a modelling approach that reflects this belief is preferable. \emph{"In this situation, not all data sources would be equally relevant, since adult data would be less relevant than paediatric data, and so a modelling approach that reflects this belief would be preferable."} Further, it recommends closer consideration when paediatric patients are included in a single trial with adults: \emph{"The appropriateness of an assumption of equally relevant data from adults and adolescents would be informed by the specific extrapolation concept and plan, and the most appropriate type of relationship should be justified and discussed with the Agency."} \cite{FDA_Bayes}.

Existing borrowing methods differ in how they handle multiple sources. MAP priors and the EXNEX framework treat all sources symmetrically: they are order-invariant and do not distinguish between adjacent and distant populations, so that the biological or temporal proximity inherent in sequential programmes is not reflected in the borrowing structure \cite{Schmidli, EXNEX}. Commensurate priors and power priors address borrowing from a single historical source \cite{Hobbs, IbrahimPowerTheory}. To our knowledge, no current framework enforces directed, adjacent-only borrowing with multiplicative weight decay across a pre-specified programme order.

This paper introduces a methodological framework that uses data-driven mixture weights to moderate borrowing when discordance is detected, while supporting prospective planning across the clinical development by pre-specifying borrowing weights and sample sizes and evaluating a comprehensive set of scenarios for the true treatment effect. By aligning sequential decision-making strategies with commonly used Go/No-Go rules \cite{lalonde}, drug developers can anticipate and manage Type I error rates, improve study assurance, and make stage-wise decisions in a coherent manner.


The remainder of this paper is organised as follows. Section 2 presents the methodological framework, including notation, formal specification of the RBSB approach, and definitions of operating characteristics for sequential decision-making. Section 3 describes the simulation study design, comparing RBSB with standalone analyses, adjacent single-source borrowing, and full pooling across nine treatment-effect trajectories. Section 4 presents simulation results, demonstrating RBSB's control of false positives whilst preserving efficiency gains. Section 5 applies the method to the START trial of budesonide (Pulmicort) in mild asthma patients 5-66 years of age, illustrating three-stage borrowing across adult, adolescent, and paediatric populations. Section 6 discusses the framework's advantages, limitations, and extensions to broader evidence synthesis settings.

\section{Methodology}
Paediatric and regional studies often warrant extrapolation from similar populations because they can be difficult to recruit. Information from neighbouring populations is often relevant to the new population, yet the new population may be distinct from the previously investigated ones. Many methods have been proposed for bridging a single source of information to a new trial, such as robust mixture priors (see Burman et al.\cite{CAFFE} and Best et al.\cite{Best}), commensurability priors, and power priors (see Viele et al. \cite{Viele} for an overview). Methods for borrowing from multiple historical sources include the meta-analytic predictive (MAP) prior \cite{Schmidli}. For methods that borrow from many populations, effects are often assumed to arise from the same distribution (akin to a mixed-effects model), limiting varying degrees of plausible exchangeability.

We propose Robust Bayesian Sequential Borrowing (RBSB), which in each study constructs a robust mixture prior comprising (i) an informative component that aggregates adjacent, previously observed evidence along the pre-specified sequence, and (ii) a vague unit-information component that safeguards against prior-data conflict. Borrowing is strictly path-dependent, contributions from more distant studies are down weighted through data-updated posterior weights $w_j^\star$, while the prior weight $w_j \in [0,1]$ governs the intended degree of borrowing at step $j$. This permits upstream evidence to contribute via admissible adjacent paths, with graded discounting when conflicts arise, rather than pooling non-adjacent sources. In practice, the analyst pre-specifies the programme order, the unit-information prior $(\mu_0, s_0)$, the step-specific weights $w_j$ (constant or varying), and the success threshold $p^\star$; the posterior weight $w_j^\star$ is then updated from the data using the closed-form marginal likelihood ratio (see Section \ref{sec:form_decp}). The approach delivers transparent and robust dynamic borrowing while avoiding implausible extrapolation, with commensurate and power-prior perspectives providing complementary context \cite{Hobbs, IbrahimPowerTheory, Schmidli, Viele}.

The method accommodates sharp discontinuities in the trajectory of the parameters, for example, when a subsequent study indicates a null effect, automatically attenuating the downstream borrowing through the mixture weights updated with data \cite{Schmidli, GiveDataAChance}. Inference is inherently path dependent: the order in which evidence is observed determines the admissible adjacent paths and influences posterior weights, in contrast to symmetric pooling approaches. The framework integrates naturally with common decision-making schemes, including Go/No-Go rules, group sequential designs, and routine sequential decisions on whether to initiate the next study, allowing prospective evaluation and control of Type I error, power, and assurance \cite{best2024beyond, ohagan}.


The first step in RBSB is to specify an initial prior for the first study. For the treatment effect, this is typically a unit-information Normal prior, $\pi(\mu_1)=\mathcal{N}(\mu_0, s_0)$, where $\mu_0$ encodes the benefit-direction convention, and $s_0$ sets the unit-information scale \cite{Kass1995}. For subsequent studies, the analyst prospectively specifies a step-specific robustification weight $w_j \in [0,1]$ that governs the borrowing of adjacent previously observed evidence. After observing sufficient statistics $\theta_j=(\hat{\mu}_j,\hat{s}_j)$, the prior is updated to a posterior Normal mixture whose components correspond to admissible adjacent paths; the parameters of each component are precision-weighted partial pools of adjacent studies, and the effective weights are moderated by the data via $w_j^\star$ (see Section~\ref{sec:form_decp} for the closed-form update). This contrasts with standard hierarchical random-effects models, which pool all sources simultaneously and symmetrically, treating exchangeability as order-invariant and ignoring temporal or biological ordering \cite{Hobbs, Viele, Schmidli, Gelman2013BDA}.

To our knowledge, the explicit, sequential, adjacent-only borrowing scheme has been little explored; we provide a formal framework with operating characteristics and decision rules tailored to multi-population programmes.

\subsection{Notation, preliminaries and assumptions}
Here we summarise the notation used in the following sections.

\begin{center}
\begin{tabular}{ll}
$\Psi$      & Type I error \\
$\Upsilon$  & Power \\
$\mu$       & True treatment effect \\
$\Omega$    & Assurance \\
$s$         & Standard error of treatment estimate \\
$\sigma$    & Standard deviation of outcome \\
$K$         & Number of studies \\
$n$         & Sample size \\
\end{tabular}
\end{center}

For convenience, define $\theta_i = (\hat{\mu}_i, \hat{s}_i)$, where $\hat{\mu}_i$ is the estimated treatment effect and $\hat{s}_i$ its standard error (SE). For integers $i \le j$, the ordered collection from studies $i$ through $j$ is $\theta_{i:j} = (\theta_i, \theta_{i+1}, \ldots, \theta_j)$. When including the unit-information prior as a pseudo-study, set $\theta_0 = (\mu_0, s_0)$, with $s_0$ specifying the unit-information scale.


For clarity, we parametrise Normal distributions as $\mathcal{N}(\text{mean}, \text{standard deviation})$. The study-level estimator of the treatment effect is assumed to follow a large-sample Normal approximation. Specifically, for study $i$, 
$$\hat{\mu}_i \mid \mu_i \sim \mathcal{N}(\mu_i, s_i)$$

Consequently, the pair $(\hat{\mu}_i, \hat{s}_i)$ is treated as an asymptotically sufficient summary of the evidence for $\mu_i$. This approximation is typically reasonable for continuous endpoints and for suitable transformed parameters for other endpoint types (e.g. log-odds for binary outcomes, log-hazard ratios for time-to-event outcomes, and log-rates for recurrent events or counts) \cite{Schmidli, Viele}. In small samples or in extreme-probability scenarios (e.g., very low or very high event rates), the adequacy of the Normal approximation should be checked, and, if necessary, alternative modelling should be used.

\subsubsection{Borrowing weights and unit-information prior\label{ref:setw}}


The vague component is a unit-information Normal prior, $\mathcal{N}(\mu_0, s_0)$, with $\mu_0 = 0$ (no treatment effect) and $s_0$ chosen to reflect the information of one observation on the parameter scale \cite{Kass1995, Schmidli}. The choice of $s_0$ materially affects the posterior mixture: a more diffuse prior (larger $s_0$) lowers its marginal likelihood and posterior weight, increasing effective borrowing; conversely, a tighter prior (smaller $s_0$) increases its marginal likelihood and reduces borrowing \cite{Crackle2025, Schmidli}.

Let $w_i \in [0,1]$ denote the step-specific borrowing weight for the study $i$, which determines the proportion of prior mass assigned to the informative component (mixture over admissible adjacent paths) versus the vague unit-information component \cite{Schmidli, Viele}. By convention, $w_1 = 0$ (no borrowing in the first study). For $i \ge 2$, $w_i$ is a fixed, pre-specified hyperparameter reflecting the anticipated exchangeability between study $i$ and its immediately preceding evidence. Each $w_i$ pertains only to its own step and is independent of other steps; the updated posterior weight $w_i^\star$ obtained after observing $\theta_i$ is not carried forward as the prior for $w_{i+1}$ \cite{wandel2022mixture}. We focus on fixed weights to ensure transparency and stable operating characteristics. A straightforward approach is to fix the borrowing weight at each step, for example, setting $w_{i+1} = C_{i+1}$, where $C_{i+1}$ reflects the expected degree of exchangeability a priori for the adjacent link in the study $i+1$. This is computationally efficient and easy to implement \cite{Schmidli, Viele}.


Alternatively, one may specify an uncertain prior for the weight, commonly a Beta distribution: 
$$ w_{i+1} \sim \mathrm{Beta}(A, B),$$ 
which allows uncertainty quantification and allows the data to moderate effective borrowing through the posterior for $w_{i+1}$. This introduces additional complexity and may suffer from weak identifiability when the data do not clearly distinguish the informative and vague components; careful prior calibration and numerically stable implementation are recommended \cite{wandel2022mixture}. 

In this paper, we focus on fixed step-specific borrowing weights because they are widely used in practice and allow for a straightforward computation of updated posterior weights \cite{Schmidli, Viele}. Each weight $w_i$ is set prospectively; after observing $\theta_i$, effective borrowing is moderated by the data through $w_i^\star$, computed from Normal marginal likelihoods (see Section~\ref{sec:form_decp}). The weights are defined independently across the steps; the choice of $w_i$ should be guided by the expected exchangeability and empirical performance of the trial simulations and, where appropriate, the elicitation of experts. Although we focus on prospectively specified borrowing weights between studies, one could in principle re-calibrate the weights after each analysis to re-assess exchangeability and re-optimise operating characteristics. However, such adaptive re-calibration conditions the design on the observed evidence rather than the pre-specified design assumptions, which complicates the evaluation of multi-period Type I error and power because the uncertainty in the treatment effect has been partially resolved by the data used to set the weights.

\subsection{Formal Description}\label{sec:form_decp}

Having established the notation and assumptions, we now formally specify the RBSB prior construction and updating scheme for a sequential programme of K studies.

For the initial study, we specify a normal unit-information prior for the treatment effect, represented as the pseudo-study $\theta_0 = (\mu_0, s_0)$. The setting $\mu_0=0$ encodes the no-effect convention, while $s_0$ is chosen to reflect approximately one observation’s worth of information on the parameter scale \cite{Kass1995}. For a continuous mean-difference endpoint, a common choice is $s_0=\sigma$; for log-scale parameters (e.g., log-odds, log-hazard), $s_0$ should be specified on the corresponding scale \cite{Schmidli, PredConsistESS}. This prior pertains to the treatment effect and does not preclude separate priors for nuisance parameters (e.g., the control-arm rate). 
\begin{equation} 
\pi(\mu_1) = \mathcal{N}(\mu_0, s_0). 
\end{equation}

The remaining studies use a robust mixture prior \cite{Schmidli, best2024beyond, Best} that combines adjacent previously observed evidence with a unit-information component. For example, after observing study 1, the prior for study 2 is 
$$\pi(\mu_2 \mid \theta_{1}) = w_2\cdot\mathcal{N}(\hat{\mu}_1, \hat{s}_1) + (1 - w_2)\cdot\mathcal{N}(\mu_0, s_0)$$
In general, the prior for the study $j+1$ is a robust mixture over all admissible adjacent-only paths $\mathcal{P}_{1:j}$ and a unit-information component:
\begin{equation}\label{eq:mixture_prior}
       \pi(\mu_{j+1} \mid \theta_{1:j})
\;=\;
w_{j+1} \cdot \sum_{S \in \mathcal{P}_{1:j}} \mathcal{W}_{S}\, \mathcal{N}(\mu_{S}, s_{S})
\;+\;
(1 - w_{j+1})\cdot \mathcal{N}(\mu_{0}, s_{0})
\end{equation}
where $\mathcal{W}_{S} \geq 0$ are path weights summing to one across $\mathcal{P}_{1:j}$ (defined below), and $(\mu_S, s_S)$ are the precision-weighted pooled mean and standard deviation for the path $S$ (defined below). Each component of the mixture corresponds to a contiguous path; non-adjacent pooling is excluded by design.

Each path is defined as a contiguous sequence of observed studies from the index $i$ to $j$. To avoid double counting, the pseudo-study for the unit-information prior is not included in the paths; it enters the robust mixture as the separate vague component \cite{Schmidli, Viele}. We write 

\begin{equation} 
S_{i:j} = \{i,i+1,\ldots,j\}, \quad 1\leq i \leq j \leq k, 
\end{equation} 

and the set of all admissible adjacent-only paths up to step $j$ is 

\begin{equation} 
\mathcal{P}_{1:j} = \{ S_{i:j} : 1 \leq i \leq j\}
\end{equation} 

\begin{figure}
    \centering
    \includegraphics[width=0.75\linewidth]{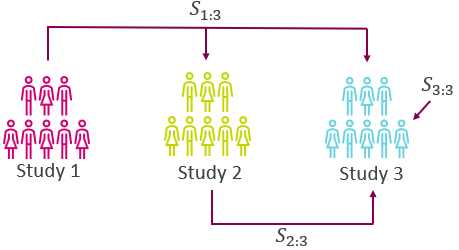}
     \caption{\emph{Illustration of paths for three studies (with information borrowing travelling from left to right).}}
    \label{fig:paths_illu}
\end{figure}

Non-adjacent paths (for example, $\{1,3\}$) are prohibited by design, ensuring that information flows only through contiguous links in the pre-specified order, as illustrated in \ref{fig:paths_illu}. 

The mean $\mu_S$ and standard deviation $s_S$ for any admissible path $S \in \mathcal{P}_{1:j}$ are obtained by precision-weighted pooling of the unit-information prior $(\mu_0, s_0)$ with the studies in $S$. Letting $\hat{\tau}_\ell = 1/\hat{s}_\ell^2$ denote the precision of study $\ell$ and $\tau_0 = 1/s_0^2$ the precision of the unit-information prior:

\begin{align*}
\hat{\tau}_S &= \tau_0 + \sum_{\ell \in S} \hat{\tau}_\ell,\\[4pt]
s_S^2 &= \frac{1}{\hat{\tau}_S}, \qquad
\mu_S = \frac{\tau_0 \mu_0 + \sum_{\ell \in S} \hat{\tau}_\ell \hat\mu_\ell}{\hat{\tau}_S}.
\end{align*}

The posterior path weights are defined recursively with the convention $w_{1}^{\star} := 0$:

\begin{equation} 
\mathcal{W}_{S_{i:j}} =(1-w^\star_i) \cdot \left(\prod_{\ell=i+1}^{j} \ w^\star_\ell\right), \quad i = 1,\ldots,j
\end{equation} 
where the product is empty (and hence equal to 1) when $i = j$, so that the singleton path $S_j = \{j\}$ has weight $\mathcal{W}_{S_j} = 1 - w^\star_j$. Here, $w_{i}^{\star} \in [0, 1]$ denotes the updated posterior mixture weight at step $i$. These weights form a properly normalised finite mixture: $\sum_{S \in \mathcal{P}_{1:j}} \mathcal{W}_{S} = 1$. See the Lemma in the Appendix~\ref{app.proof}.



In the robust mixture prior at step $j$, the posterior mixture weight $w_j^\star$ after observing $\theta_j$ is given by the following:
\[
w_j^\star
=
\frac{\, w_j \cdot m_{\text{inf}}(\theta_j) \,}
     {\, w_j \cdot m_{\text{inf}}(\theta_j) \;+\; (1 - w_j) \cdot m_{\text{vag}}(\theta_j) \,}.
\]
Here, \(m_{\text{inf}}(\theta_j)\) and \(m_{\text{vag}}(\theta_j)\) are the marginal predictive likelihoods of the observed estimator $\hat{\mu}_j$ under the informative and vague components, respectively \cite{Best}.

Letting \(\phi(x; m, s)\) denote the Normal density with mean \(m\) and standard deviation \(s\), the marginal likelihoods at the analysis step \(j\) (with embedded pseudo-study, adjacent-only paths and previously defined path weights \(\mathcal{W}_S\)) are:
\[
m_{\text{inf}}(\theta_j)
=
\sum_{S \in \mathcal{P}_{1:j-1}} \mathcal{W}_{S} \;
\phi\!\left( \hat{\mu}_j;\, \mu_{S},\, \sqrt{\, s_{S}^{2} + \hat{s}_j^{2} \,} \right),
\]
\[
m_{\text{vag}}(\theta_j)
=
\phi\!\left( \hat{\mu}_j;\, \mu_0,\, \sqrt{\, s_0^{2} + \hat{s}_j^{2} \,} \right),
\]


After observing study $i+1$, the posterior updates the robust prior to a finite mixture over adjacent-only paths through the accrued evidence: 

\begin{equation}\label{eq.bb} 
\pi(\mu_{i+1} \mid \theta_{1:(i+1)}) = \sum_{S\in \mathcal{P}_{1:(i+1)}} \mathcal{W}_{S}\cdot \mathcal{N}(\mu_{S}, s_{S}). 
\end{equation} 
The fitting of the model proceeds sequentially: in the analysis step $i$ (the analysis of the study $i$ using the prior constructed from the studies $1$ through $i-1$) we update a single borrowing hyperparameter $w_i$ to its posterior weight driven by data $w_i^\star$. We do not refit mixture weights from earlier steps; rather, path weights propagate multiplicatively through the sequence. This contrasts with simultaneous, single-period mixture construction at one time point (e.g., robust MAP prior assembly) where all sources are fitted together \cite{Schmidli}. Under the multiplicative path-weight recursion, the contribution of any distant path can only remain unchanged (if the new weight equals one) or decrease as the programme advances, it cannot increase.

Algorithm~\ref{alg:RBSB_step} summarises the RBSB analysis of study $j$ ($j \ge 2$) given the previously analysed studies $1,\ldots,j{-}1$.

\begin{algorithm}
\caption{RBSB: analysis of study $j$ in a sequence of $K$ studies ($j \ge 2$)}
\label{alg:RBSB_step}
\begin{flushleft}
\textbf{Input:} $\theta_{1:j}=\{(\hat\mu_\ell,\hat s_\ell)\}_{\ell=1}^{j}$;\; $(\mu_0, s_0)$;\; $w_j$;\; $\{w^\star_\ell\}_{\ell=1}^{j-1}$ (with $w^\star_1:=0$);\; $p^\star$\\[4pt]
\textbf{1.}\; \textbf{Paths \& pooled parameters.}\;
$\mathcal{P}_{1:j-1} \leftarrow \{S_{i:j-1}: 1\le i\le j{-}1\}$.\;
For each $S\!\in\!\mathcal{P}_{1:j-1}$:
$\;\hat{\tau}_S \leftarrow \tau_0{+}\textstyle\sum_{\ell\in S}\hat\tau_\ell$,\;
$s_S^2 \leftarrow 1/\hat{\tau}_S$,\;
$\mu_S \leftarrow (\tau_0\mu_0{+}\textstyle\sum_{\ell\in S}\hat\tau_\ell\hat\mu_\ell)/\hat{\tau}_S$.\\[3pt]
\textbf{2.}\; \textbf{Path weights.}\;
$\mathcal{W}_{S_{i:j-1}} \leftarrow (1{-}w^\star_i)\prod_{\ell=i+1}^{j-1}w^\star_\ell$,\; $i=1,\ldots,j{-}1$.\\[3pt]
\textbf{3.}\; \textbf{Robust mixture prior.}\;
$\pi(\mu_j) \leftarrow w_j\!\sum_{S\in\mathcal{P}_{1:j-1}}\!\mathcal{W}_S\,\mathcal{N}(\mu_S,s_S) \;+\; (1{-}w_j)\,\mathcal{N}(\mu_0,s_0)$.\\[3pt]
\textbf{4.}\; \textbf{Observe} $\theta_j$ \textbf{and update weight.}\;
$w^\star_j \leftarrow \dfrac{w_j\, m_{\mathrm{inf}}}{w_j\, m_{\mathrm{inf}} + (1{-}w_j)\, m_{\mathrm{vag}}}$,\;
where $m_{\mathrm{inf}}\!=\!\sum_S \mathcal{W}_S\,\phi(\hat\mu_j;\mu_S,\sqrt{s_S^2{+}\hat{s}_j^2})$,\; $m_{\mathrm{vag}}\!=\!\phi(\hat\mu_j;\mu_0,\sqrt{s_0^2{+}\hat{s}_j^2})$.\\[3pt]
\textbf{5.}\; \textbf{Posterior.}\;
$\pi(\mu_j\!\mid\!\theta_{1:j}) \leftarrow \sum_{S\in\mathcal{P}_{1:j}} \mathcal{W}_S\,\mathcal{N}(\mu_S,s_S)$,\;
with $\mathcal{P}_{1:j} = \{S_{i:j}: 1\le i\le j\}$ and updated path weights using $w^\star_j$.\\[3pt]
\textbf{6.}\; \textbf{Decision:} declare success if $\Pr(\mu_j>0\mid\theta_{1:j})>p^\star$.
\end{flushleft}
\end{algorithm}

\subsection{Decision Criteria and operating characteristics}\label{sec:dec_criteria}
To evaluate and compare borrowing designs prospectively, it is necessary to define formal operating characteristics such as Type~I error, power, and assurance, that summarise how a given decision rule performs under specified true-effect trajectories. Without such definitions, it is not possible to assess whether a design adequately controls false positives or achieves sufficient power across the sequence of studies. 

A typical clinical trial decision rule declares success if the treatment effect is significant at the one-sided 2.5\% level. When there is robust evidence from related populations, a less stringent threshold can sometimes be justified. In this framework, we consistently define the study success in all analyses as 
$$\Pr(\mu_i > 0 \mid \theta_{1:i}) > p^\star,$$
with $p^\star=0.975$ unless otherwise specified. Unlike traditional frequentist criteria, this Bayesian approach leads to operating characteristics, such as power and Type~I error, that depend on the prior, which may differ between studies. The appropriate metric depends on what is known at the time of planning: marginal metrics apply when upstream evidence is already observed and enters the analysis as fixed input; conditional and joint metrics are suited to prospective programme planning when upstream results are still to be collected and remain stochastic. Assurance, denoted $\Omega$, extends this further by integrating power over a prior distribution for the true treatment effect rather than conditioning on a fixed value \cite{ohagan}; see Appendix~\ref{app.assurance}. We therefore distinguish marginal, conditional, and joint versions of these concepts below.

\subsubsection{Marginal Type I Error \& Power}\label{sec.reg_metric}

Let $D_j$ denote the study-level decision indicator under a "higher is better" benefit-direction convention: 
\begin{equation}
    D_j = \mathbbm{I}\left\{ \Pr\big(\mu_j > 0 \big| \theta_{1:j} \big) > p^\star \right\}.
\end{equation}
The marginal Type I error and the marginal power in stage $j$, conditioned on previously observed evidence but not on previous decisions, are 
\begin{equation}\label{eq:t1e} 
\Psi_j = \Pr\left(D_j = 1 \middle| \theta_{1:j-1}, \mu_{j} = 0 \right), 
\end{equation} 
\begin{equation}\label{eq:pwr} 
\Upsilon_j = \Pr\left(D_j = 1 \middle| \theta_{1:j-1}, \mu_{j} = \delta \right), 
\end{equation} 
where $\delta$ denotes a clinically relevant alternative for planning. These quantities are appropriate for stage-specific evaluation when upstream evidence is fixed (for example, adult results observed when planning adolescents).

\subsubsection{Conditional Metrics}\label{sec.condmmetric}
When progression to study $j{+}1$ is contingent on success in previous stages, conditional operating characteristics reflect performance under binding programme rules. For example, if efficacy is not demonstrated in adults and adolescents, a paediatric study may not proceed.

The conditional Type~I error and the power in stage $j$, given the success in stages $i,\ldots,j{-}1$, are 
\begin{equation}\label{eq:conditional_t1e} 
\Psi_{\mathcal{P}_{i:j}}^{(C)} = \Pr\left(D_j = 1 \middle| \bigcap_{l=i}^{j-1} D_l = 1, \theta_{1:j-1}, \mu_j = 0\right), 
\end{equation} 
\begin{equation}\label{eq:conditional_power} 
\Upsilon_{\mathcal{P}_{i:j}}^{(C)} = \Pr\left(D_j = 1 \middle| \bigcap_{l=i}^{j-1} D_l = 1, \theta_{1:j-1}, \mu_j = \delta\right). 
\end{equation} 
These metrics condition on both the accrued evidence and the prior-stage decisions; the conditioning event may reflect true or false positives/negatives and thus captures the dependence induced by sequential borrowing and programme rules. In practice, the conditioning event can have low probability in long programmes or when success probabilities are modest, yielding sparse Monte Carlo strata and inflated estimation uncertainty; it is prudent to increase simulation budgets and report simulation-based confidence intervals in such settings.

\subsubsection{Joint Metrics}\label{sec.jointmetric}
Joint metrics quantify performance across several stages simultaneously under the programme design. The joint Type I error and the joint power in          stages $i,\ldots,j$ are 
\begin{equation}\label{eq:joint_t1e} 
\Psi_{\mathcal{P}_{i:j}}^{(J)} = \Pr\left( \bigcap_{l=i}^{j} D_l = 1 \middle| \theta_{1:j-1}, \mu_j = 0\right), 
\end{equation} 
\begin{equation}\label{eq:joint_power} 
\Upsilon_{\mathcal{P}_{i:j}}^{(J)} = \Pr\left( \bigcap_{l=i}^{j} D_l = 1 \middle| \theta_{1:j-1}, \mu_j = \delta\right). 
\end{equation} 
These quantities are most relevant for seamless programmes, where new stages or populations are added only if previous evidence is successful. Joint metrics are bounded above by marginal metrics, so power and Type I error for the overall sequence will generally be lower than at individual stages.

\subsection{Evaluation metrics}\label{sec.evalmet}

In sequential programmes with borrowing, simulation scenarios specify a trajectory of true effects across stages rather than a single global hypothesis. Consequently, we report the rejection rate, defined as the Monte Carlo probability that the decision rule is triggered under a given scenario trajectory. When the true effect at a stage equals the null value, the stage rejection rate coincides with the Bayesian Type~I error for that stage; when it equals a clinically relevant alternative, it coincides with power. We report marginal, conditional and joint rejection rates as defined above, together with bias and effective sample size percentage (ESS\%), to allow direct comparison of borrowing strategies and their operating characteristics. Bias at stage~$i$ is defined as the Monte Carlo average of $\tilde{\mu}_i - \mu_i$, where $\tilde{\mu}_i$ denotes the posterior median and $\mu_i$ the true treatment effect. ESS\% at stage~$i$ is defined as $(\mathrm{ESS}_i - n_i)/\mathrm{ESS}_i$, averaged across simulations, where $\mathrm{ESS}_i$ is the effective sample size of the posterior \cite{PredConsistESS} and $n_i$ is the study sample size. It measures the proportion of the posterior information attributable to borrowed data, with $\mathrm{ESS\%} = 0$ indicating no borrowing and larger values indicating greater prior influence.

\section{Simulation setups}\label{sec.sim_setup}
We conduct a simulation study to evaluate the operating characteristics of sequential, adjacent-only Bayesian borrowing in multi-population clinical trial programmes. Specifically, we compare Robust Bayesian Sequential Borrowing (RBSB) with standalone analyses, adjacent single-source borrowing, and full pooling under prospectively defined programme orders. The primary metrics are marginal, conditional and joint Type~I error, power, bias and percentage borrowed data, assessed in scenarios that reflect gradients of similarity and potential discordance between populations. Fixed sample sizes are used to illustrate design behaviour and enable comparability between methods; in practice, these can be adapted to feasibility and programme objectives, with calibration via prospective simulations.

\subsection{Study designs to evaluate} 
We compare four analysis strategies for multi-population programmes, each reflecting distinct assumptions about exchangeability and information flow across populations.

\begin{description}

\item[Standalone] Each population is analysed independently without borrowing from historical or adjacent sources. This design serves as a reference for Type~I error control and power under minimal assumptions, ensuring that inference is driven solely by new data.

\item[Full pooling Bayesian borrowing] All available historical and concurrent data are combined via a fixed-effect pooling model under an implicit homogeneity assumption across subpopulations (i.e., effects are exchangeable); the pooled estimate is used as the informative component of a robust mixture prior with a vague unit-information component. Although this maximises borrowing, it can inflate false positives and bias estimates when heterogeneity or a conflict between prior data is present \cite{Schmidli,Viele}.

\item[Adjacent Bayesian borrowing] At each stage, a robust mixture prior is formed using only the immediately preceding study's estimate as the informative component and a unit-information component as the vague component; non-adjacent sources are excluded. This enforces proximity (e.g.\ adults $\rightarrow$ adolescents $\rightarrow$ paediatrics), reflecting stronger a priori exchangeability between nearest neighbours and mitigating implausible extrapolation across distant populations \cite{Viele,GiveDataAChance}.

\item[Robust Bayesian Sequential Borrowing (RBSB)] A sequential, adjacent-only framework that uses robust mixture priors that combine an informative component with a unit-information component to protect against prior-data conflict, see Section \ref{sec:form_decp}. Borrowing is path-dependent and dynamically attenuated via posterior weights derived from predictive marginal likelihoods; path weights propagate multiplicatively across the sequence, preventing non-adjacent pooling and aligning with the prespecified programme order.

\end{description}

\subsection{Simulation Scenarios}
We model a sequence of five trials, for example, adults~$\rightarrow$ adolescents (12--17 years)~$\rightarrow$ children (6--11 years)~$\rightarrow$ children (2--5 years)~$\rightarrow$ infants ($<$2 years), assuming that adjacent populations are more similar than distant ones. Study-level treatment-effect estimators follow a Normal distribution on the analysis scale. Specifically, 
\begin{equation} 
\hat{\mu}_i \mid \mu_i \sim \mathcal{N}\left(\mu_i, \frac{\sigma}{\sqrt{n_i}} \right) 
\end{equation} 
with $\sigma$ specified on the analysis scale. We consider scenarios that reflect gradients of similarity and potential discordance throughout the sequence; these are summarised in Table~\ref{tab:scenarios}. Unless otherwise stated, sample sizes are $n_1=160$ and $n_2=n_3=n_4=n_5=60$, with target alternative $\delta=0.5$ and common standard deviation $\sigma=1$. Decisions adopt a one-sided posterior tail-probability rule.

The borrowing weights are fixed across the studies for clarity and comparability. We set $w_1=0$ (no borrowing in the first study) and $ w_i = 0.5,\quad i=2,3,4,5$. The unit‑information prior is set to $\mu_0=0$ and $s_0=1$, following the approach described in Section~\ref{ref:setw}. Using fixed weights simplifies interpretation of operating characteristics and isolates the impact of the borrowing scheme itself; weight calibration to reflect anticipated exchangeability can be considered separately via prospective simulations.

\begin{table}[ht] 
\centering 
\begin{tabular}{cl ccccc}
\toprule
& \textbf{Setting} & \textbf{Study 1} & \textbf{Study 2} & \textbf{Study 3} & \textbf{Study 4} & \textbf{Study 5} \\
\midrule
\multirow{1}{*}{{\textit{Global Null}}}
& 1 & $0$      & $0$      & $0$      & $0$      & $0$      \\
\midrule
\multirow{5}{*}{{\textit{Scenario 1}}}
& 2 & $\delta$ & $0$      & $0$      & $0$      & $0$      \\
& 3 & $\delta$ & $\delta$ & $0$      & $0$      & $0$      \\
& 4 & $\delta$ & $\delta$ & $\delta$ & $0$      & $0$      \\
& 5 & $\delta$ & $\delta$ & $\delta$ & $\delta$ & $0$      \\
& 6 & $\delta$ & $\delta$ & $\delta$ & $\delta$ & $\delta$ \\
\midrule
\multirow{3}{*}{{\textit{Scenario 2}}}
& 7 & $\delta$ & $\delta$ & $0.6\!\cdot\! \delta$ & $0$  & $0$  \\
& 8 & $\delta$ & $\delta$ & $0.6\!\cdot\!\delta$ & $0.6\!\cdot\!\delta$ & $0$  \\
& 9 & $\delta$ & $\delta$ & $0.6\!\cdot\!\delta$ & $0.6\!\cdot\!\delta$ & $0.6\!\cdot\!\delta$ \\
\bottomrule
\end{tabular}
\caption{Overview of simulation scenarios and true treatment effects ($\mu_i$) by study. Settings~1--6 assume a constant effect across active stages; Settings~7--9 allow the effect to attenuate to $0.6\!\cdot\!\delta$ in later stages.} \label{tab:scenarios}
\end{table}

\section{Simulation results}\label{sec.simres}

In this section, we present results for the simulation set-up described in Section \ref{sec.sim_setup}. We used fixed sample sizes and fixed borrowing weights to illustrate design behaviour in nine treatment-effect trajectories (Table \ref{tab:scenarios}). Rejection rates are reported under the one-sided posterior tail-probability rule $\Pr(\mu_i>0 \mid \theta_{1:i}) > 0.975$, where a rejection at a stage with a true null corresponds to a Bayesian Type I error and a rejection at a stage with the clinically relevant alternative corresponds to power. We summarise marginal, conditional, and joint rejection rates as defined in Section \ref{sec:dec_criteria}, together with bias and effective sample size percentage (ESS\%), to compare standalone analysis, full pooling, adjacent-only borrowing, and RBSB.

Across all settings, RBSB consistently moderates borrowing when discordance arises between adjacent populations while preserving efficiency when effects are aligned. When all true effects are null (Setting 1), all methods exhibit low stage-wise rejection rates, with RBSB similar to standalone and below full pooling at later stages, reflecting improved Type I error control relative to full pooling. When early stages have non-zero effects and later stages are null (Settings 2–5), full pooling inflates Type I error in downstream null stages, while adjacent-only borrowing reduces inflation, but can under-use compatible upstream information. RBSB achieves higher power than adjacent-only borrowing in aligned stages and markedly lower false positives than full pooling in discordant stages. When all stages have non-zero effects (Setting 6), RBSB retains most of the power gains of pooling while keeping bias and ESS\% at moderate levels. Settings 7--9 introduce reduced effects ($0.6\!\cdot\!\delta$) in later stages, looking at the operational characteristics when partial efficacy persists downstream. RBSB borrows more than adjacent-only borrowing in the stages where the attenuated effect is present, improving power, yet decreases borrowing more strongly than full pooling in the null stages that follow, resulting in lower Type~I error inflation.

\begin{figure}
    \centering
    \includegraphics[width=\linewidth]{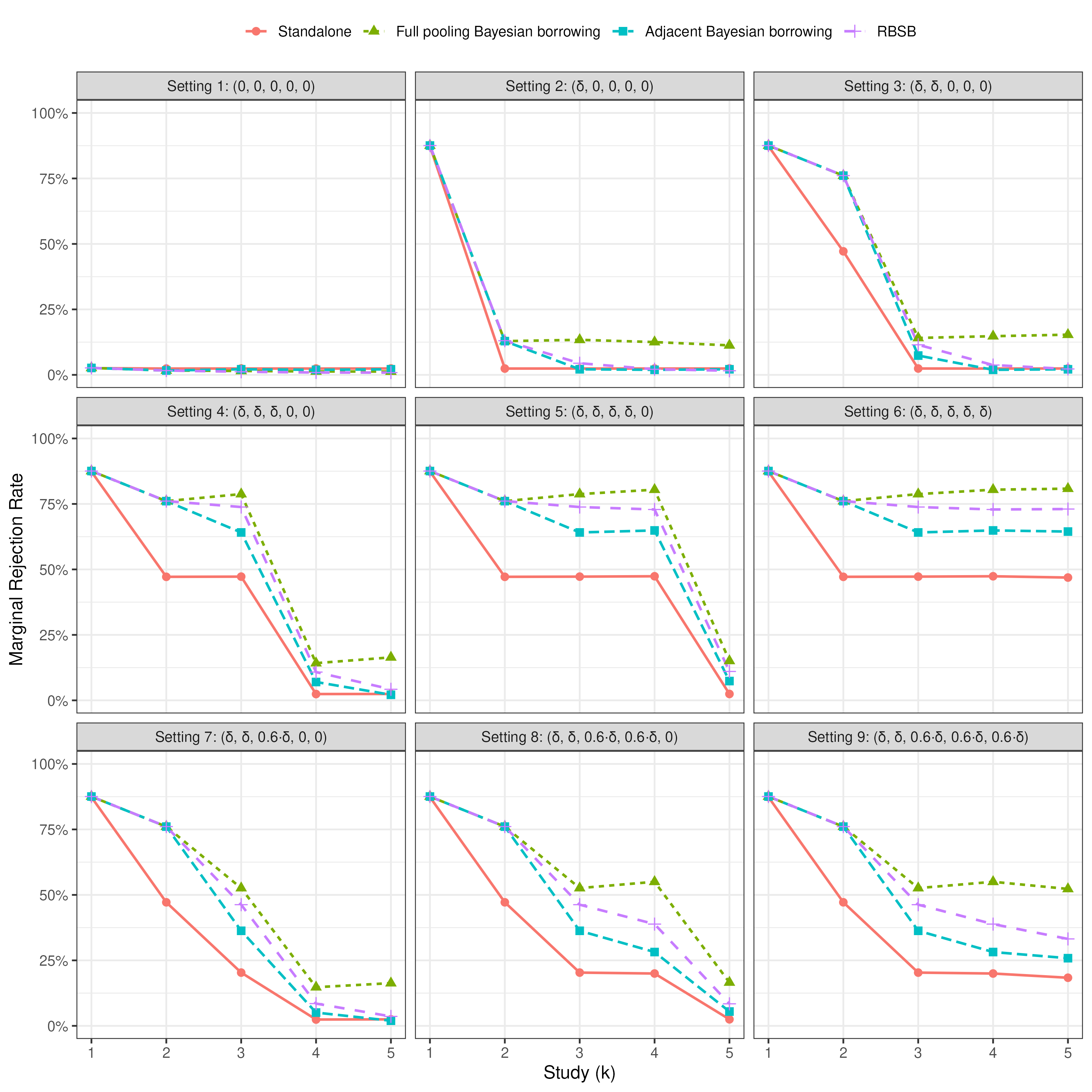}
    \caption{\emph{Simulation results for marginal evaluation metrics (see Section~\ref{sec.evalmet} and Table~\ref{tab:marginal}). The difference between Settings~1 and~2 is the true treatment effect in Study~1 (0 vs $\delta$).}}
    \label{fig:margrates}
\end{figure}

Marginal results are plotted in Figure \ref{fig:margrates} and Table \ref{tab:marginal} summarises marginal rejection rates by stage. In Setting 1 (all null), RBSB's marginal rejection remains close to the nominal level across stages and below the inflation seen with the full pooling in later stages. In Settings with early non-zero effects followed by nulls (for example, Settings 2–4), full pooling shows clear inflation in the null stages, while RBSB reduces this inflation substantially. In aligned-effect settings (Settings 5–6), RBSB's marginal power closely tracks full pooling and exceeds adjacent-only borrowing, reflecting effective but disciplined borrowing. In settings with attenuated-effect (Settings 7-8), where later stages carry a reduced effect of $0.6\!\cdot\!\delta$, RBSB achieves a marginal power intermediate between full pooling and borrowing only adjacent in the active stages, while the marginal Type~I error in subsequent null stages remains lower than in full pooling.

\begin{figure}
    \centering
    \includegraphics[width=\linewidth]{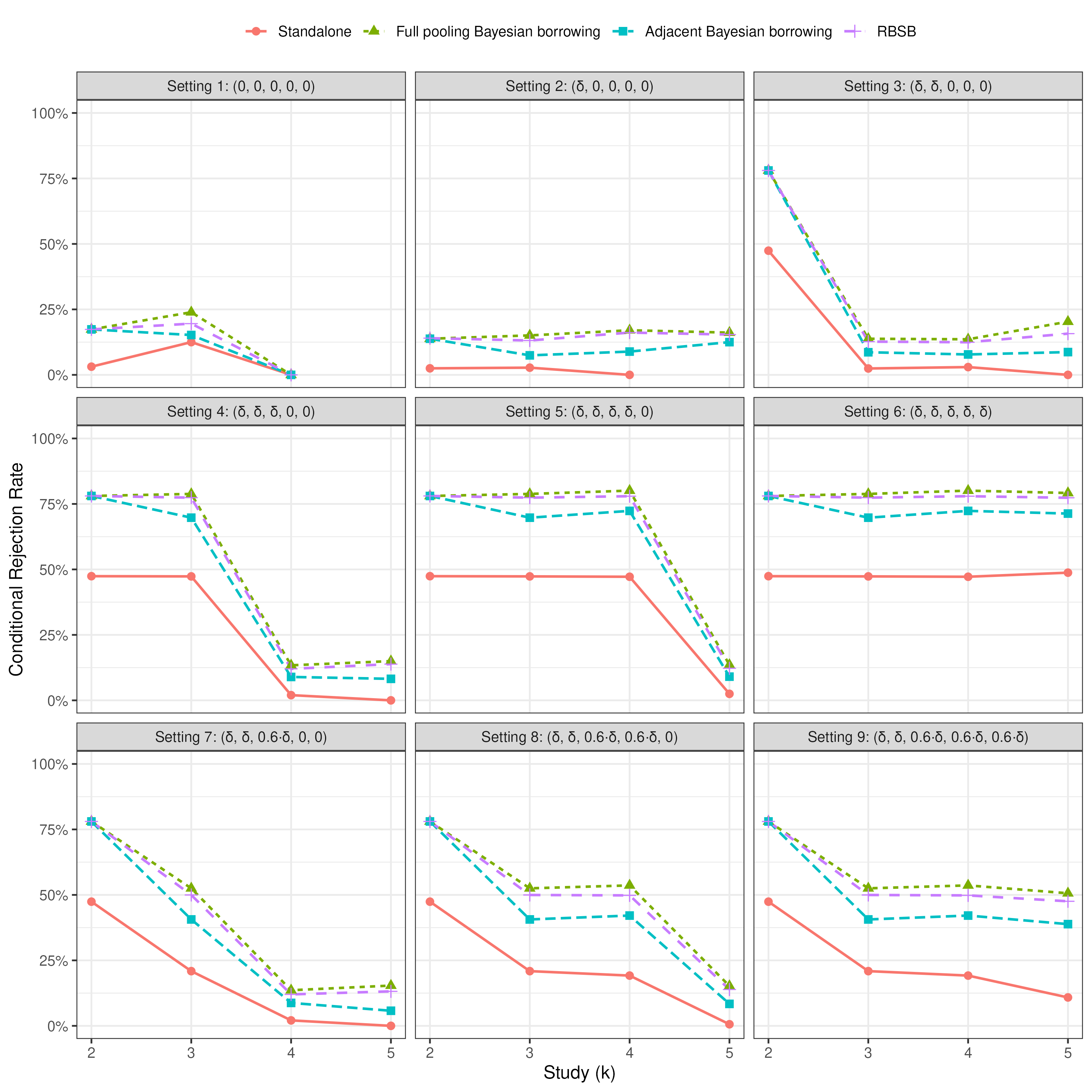}
    \caption{\emph{Simulation results for conditional evaluation metrics (see Section~\ref{sec.condmmetric} and Table~\ref{tab:cond}). The difference between Settings~1 and~2 is the true treatment effect in Study~1 (0 vs $\delta$). Missing values (e.g., Trial~5 in Setting~1) occur when no simulation replicate reaches the required conditioning stage.}}
    \label{fig:condrates}
\end{figure}

Conditional rejection rates (Figure \ref{fig:condrates} and Table \ref{tab:cond}), which are conditional on success in previous stages, accentuate differences between designs. Full pooling shows the highest conditional Type I error in later null stages, driven by selection on earlier success and continued strong borrowing. Adjacent-only borrowing reduces this conditional inflation, but at the cost of lower conditional power when effects remain aligned. RBSB shows an improved calibration: conditional power remains high when effects are consistent, while conditional Type I error is attenuated by data-driven down weighting when conflicts arise.
Note that some entries are reported as NA in Figure \ref{fig:condrates} and Table \ref{tab:cond}. This occurs because in certain scenarios no simulations progress to the future stage required for conditioning, leaving the conditional probability undefined. This underscores the difficulty of interpreting conditional metrics in distant future stages with low progression probability.

\begin{figure}
    \centering
    \includegraphics[width=\linewidth]{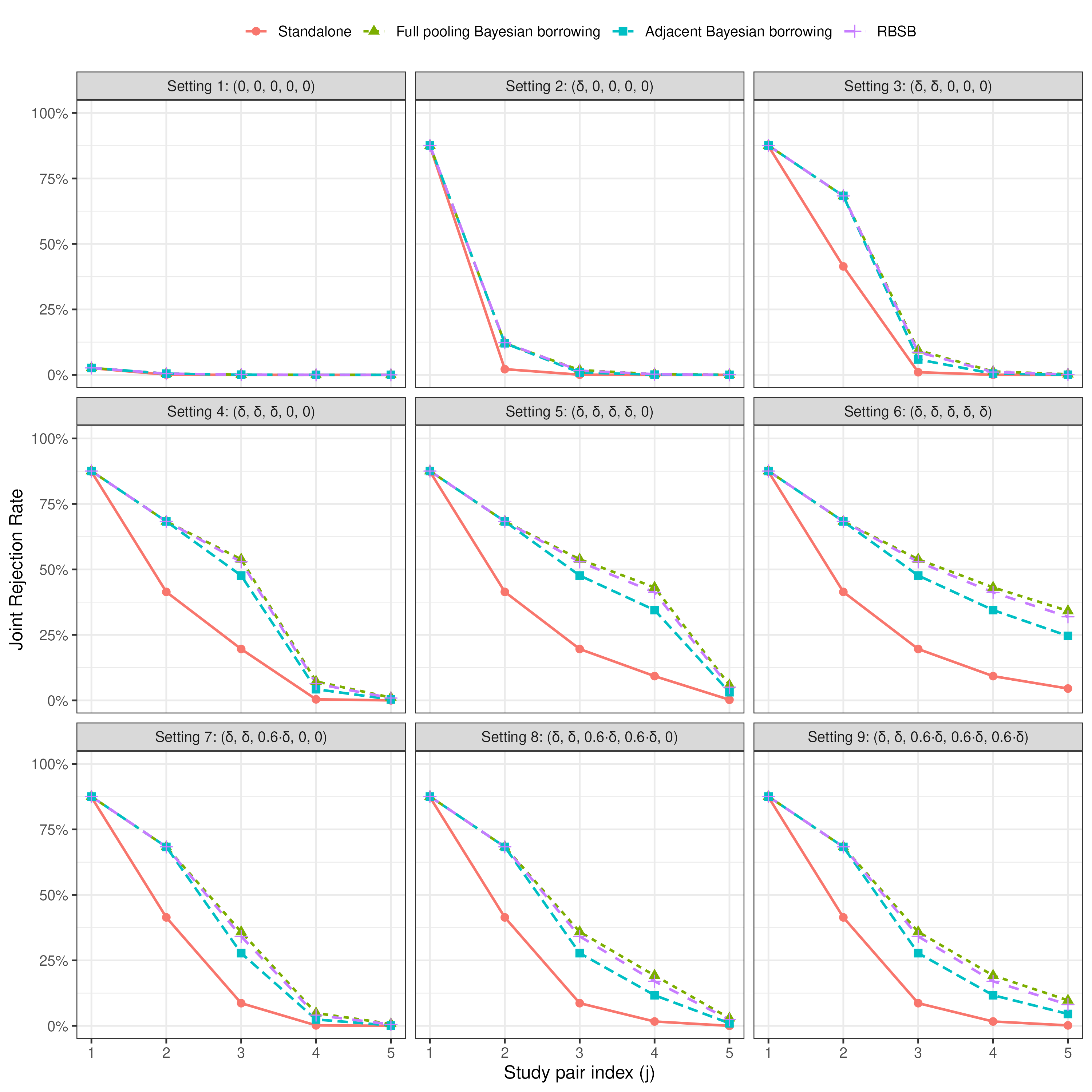}
    \caption{\emph{Simulation results for joint evaluation metrics (see Section~\ref{sec.evalmet} and Table~\ref{tab:joint}).}}
    \label{fig:jointrates}
\end{figure}

The joint rejection rates in multiple stages (Figure \ref{fig:jointrates} and Table \ref{tab:joint}) show the cumulative consequences of each borrowing strategy. Full pooling yields the largest joint rejection probabilities when effects are aligned, but it also produces the highest joint false-positive rates when later stages are null. RBSB maintains strong joint power in aligned settings and reduces joint false positives compared to full pooling in discordant settings, consistent with its path-dependent attenuation.

\begin{figure}
    \centering
    \includegraphics[width=\linewidth]{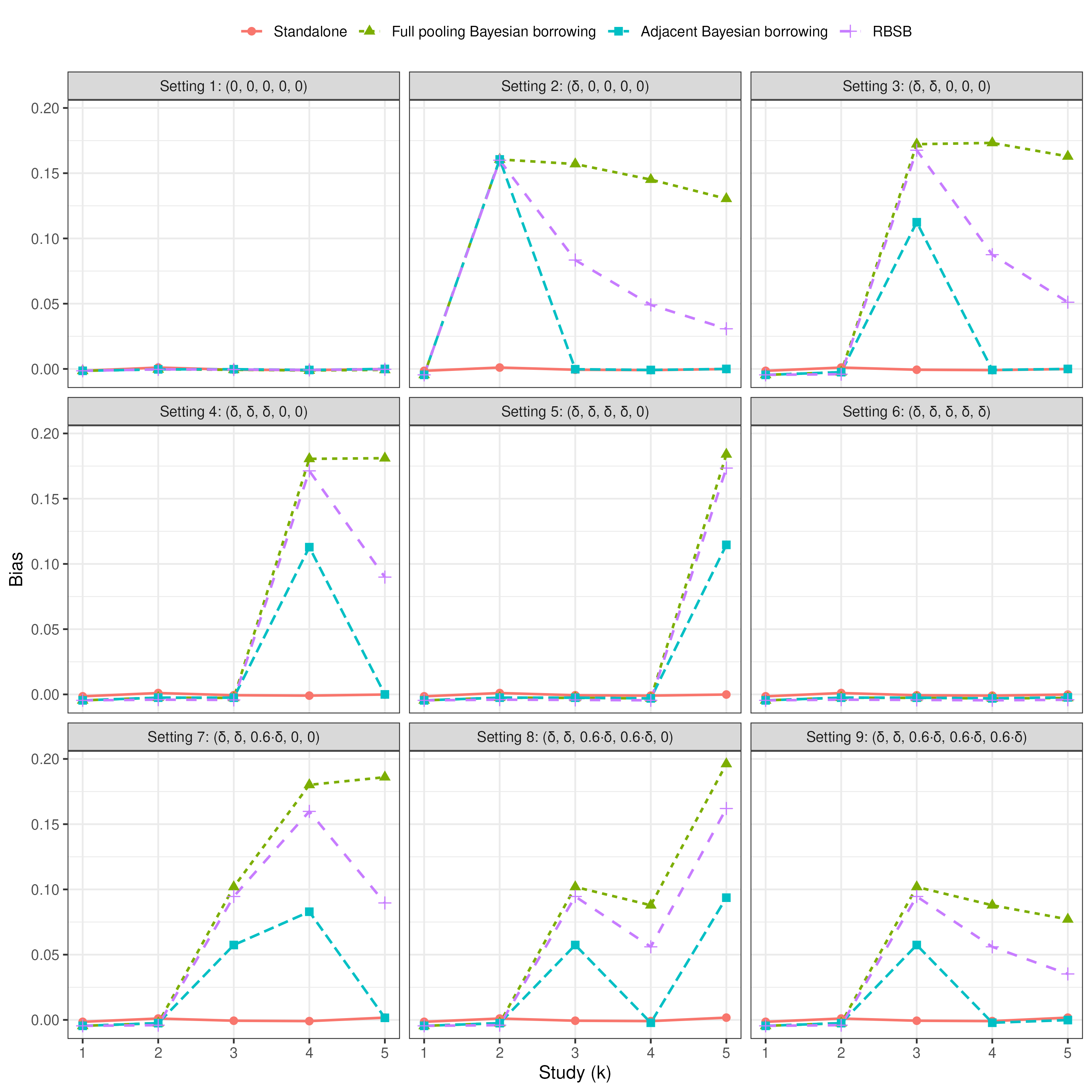}
    \caption{\emph{Simulation results for bias evaluation metrics (see Section~\ref{sec.evalmet} and Table~\ref{tab:bias}).}}
    \label{fig:Bias}
\end{figure}

\begin{figure}
    \centering
    \includegraphics[width=\linewidth]{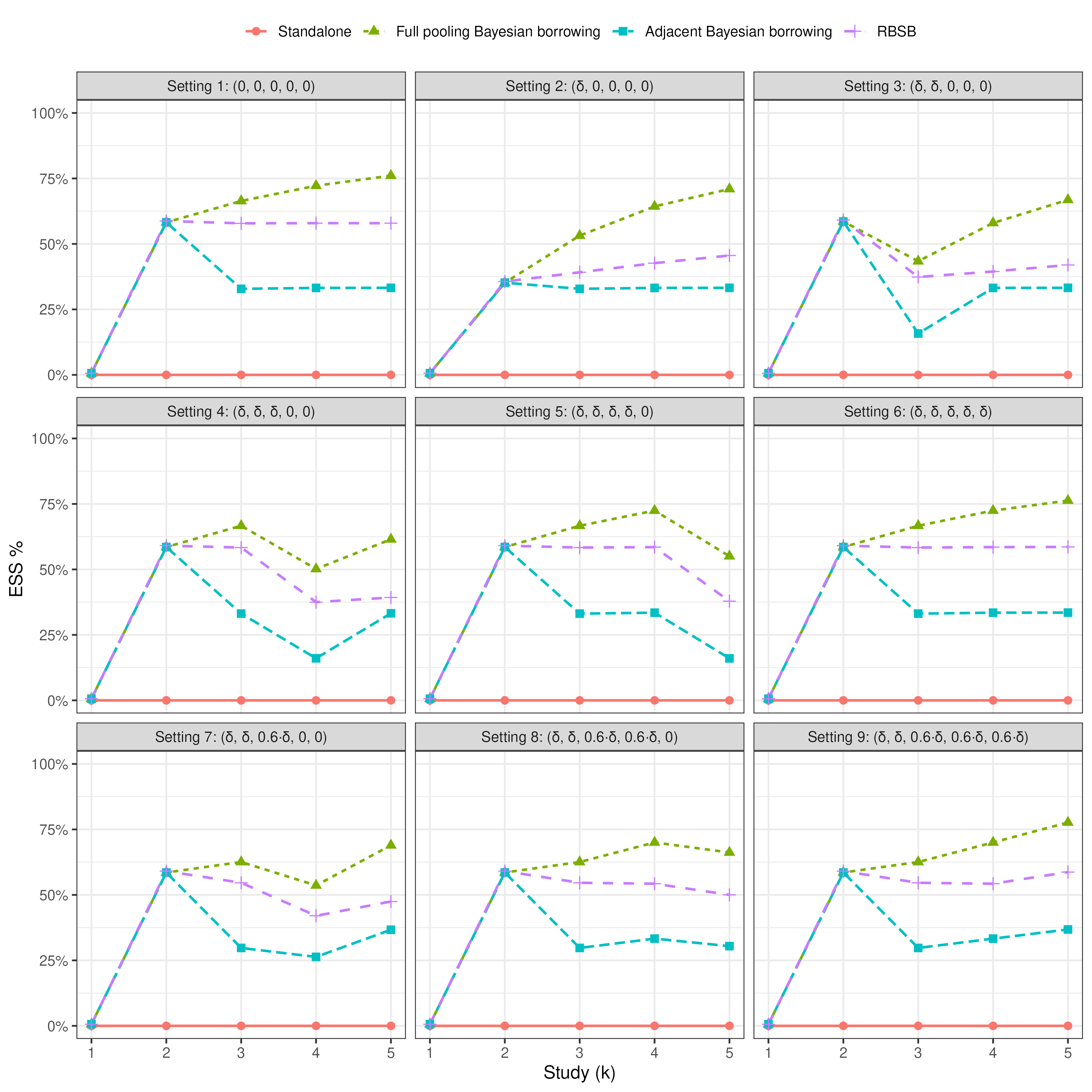}
    \caption{\emph{Simulation results for ESS evaluation metrics (see Section~\ref{sec.evalmet} and Table~\ref{tab:ess}).}}
    \label{fig:ESS}
\end{figure}

Bias and effective sample size ESS\% (Figure \ref{fig:Bias} \& \ref{fig:ESS}, Table \ref{tab:bias} \& \ref{tab:ess}) illuminate the mechanisms underlying performance. Full pooling generates the highest ESS\% and the largest bias in discordant settings due to persistent borrowing from misaligned sources. Adjacent-only borrowing keeps ESS\% low downstream, which limits bias inflation but also forfeits efficiency when evidence is compatible. RBSB borrows a moderate amount (as can be seen in ESS\%), with reduced borrowing when a prior data conflict is detected. This manages bias while preserving meaningful efficiency gains in compatible settings. This behaviour reflects the multiplicative decay of longer paths through updated mixture weights and aligns with the design intent of path-dependent borrowing.


\FloatBarrier


\FloatBarrier
\FloatBarrier

\section{Case Study}\label{sec.case_study}

In this section, we present a case study applying our proposed methodology to the START study (NCT00641914), a randomised, double‑blind, three‑year study in patients with asthma treated with budesonide (Pulmicort) or non-active treatment, completed in 2009. The trial recruited 7,241 patients with mild persistent asthma of less than two years of duration who had not previously taken glucocorticosteroids regularly. The participants were between 5 and 66 years old. All active‑arm patients aged 11 years and older received 400 $\mu$g daily, while those under 11 years received 200 $\mu$g daily \cite{PAUWELS20031071}. The study was designed to detect a 35\% reduction in the hazard of experiencing an asthma‑related event, with 95\% power at a two‑sided 5\% significance level. Although this is a highly powered study overall, subgroup analyses naturally have lower power. The results are summarised in Table~\ref{tab:start_study}; they were not reported in the main publication, but were provided by the study team.

The primary endpoint was the time to the first severe asthma-related event. The study evaluated three age strata: adults (18 years and older), adolescents (11–17 years), and paediatrics (below 11 years). Although these groups were recruited concurrently rather than sequentially, the setting is suitable for illustrating RBSB's path-dependent borrowing framework. The mode of action and exposure are plausibly similar between age groups and, as is typical in clinical trials, the paediatric and adolescent subgroups are smaller than the adult group. We therefore treat the three strata as adjacent populations in sequence, applying RBSB to demonstrate disciplined borrowing that respects biological proximity while guarding against prior-data conflict.

\begin{table}[htb] 
\centering 
\begin{tabular}{|c|c|c|} 
\hline 
Population & N & HR (95\% CI) \\
\hline 
Adult (18-66 yr) & 3970 & 0.54(0.40, 0.74) \\ 
Adolescents (11-17 yr) & 1221 & 0.56 (0.30, 1.04)\\ 
Children (5-10 yr) & 1974 & 0.60 (0.40, 0.90)\\
\hline 
\end{tabular} 
\caption{\emph{Summary statistics for the START study}} 
\label{tab:start_study} 
\end{table}

We make the following simplifying assumptions: 1:1 randomisation (as in this study) and proportional hazards within each subgroup. Under these assumptions, we use the approximation 
\begin{equation} 
\mathrm{Var}\big(\log(\mathrm{HR})\big) \approx \frac{4}{D}
\end{equation} 
where $D$ is the total number of events observed in the subgroup. This rule of thumb allows us to infer the total number of events for each subpopulation from the log‐hazard ratios and confidence intervals reported and from this calculate the effective sample size percentage (ESS) for each subgroup analysis. Consequently, we adopt a unit‑information prior $\mathcal{N}(0,2)$, centred on a hazard ratio of 1 and corresponding to the information from a single event on the log‑hazard scale.

\begin{table}[htb] \centering 
\begin{tabular}{|c|c|c|} 
\hline 
Population & N Events & Log HR (95\% CI) \\ 
\hline 
Adult (18-66 yr) & 162 & -0.62 (-0.92, -0.3) \\ 
Adolescents (11-17 yr) & 40 & -0.58 (-1.2, 0.04)\\ 
Children (5-10 yr) & 93 & -0.51 (-0.92, -0.11)\\
\hline
\end{tabular} 
\caption{\emph{Derived numbers of events and log hazard ratios.}} 
\label{tab:start_study_log} 
\end{table}


We evaluated these outcomes using the RBSB approach, assigning a fixed prior borrowing weight of $w=0.5$ to each adjacent link between neighbouring age strata. In principle, these weights should be determined prospectively based on prior information and elicited exchangeability judgements; however, in this illustrative example, we did not examine biological plausibility and instead chose a common value solely for demonstration purposes. 

\subsection{Analysis per stage}

In the following sections, we walk through the RBSB construction stage by stage, instantiating the general framework of Section~\ref{sec:form_decp} with the START trial data (Table~\ref{tab:start_study_log}).

\subsubsection{Stage 1: Adults}

The unit-information prior for the treatment effect on the log-hazard scale is
\begin{equation}
\pi(\mu_1) = \mathcal{N}(0,\, 2),
\end{equation}
centred on no treatment effect ($\mu_0=0$, i.e.\ $\mathrm{HR}=1$), with $s_0=2$ reflecting approximately one event's worth of information on the log-hazard scale.

After observing the adult summary $\theta_1 = (\hat\mu_1, \hat s_1) = (-0.62,\, 0.16)$ (see Table~\ref{tab:start_study_log}), the pooled path parameters for $S_1=\{1\}$ are obtained by precision-weighted pooling of the unit-information prior $(\mu_0=0, s_0=2)$ with study~1, as defined in Section~\ref{sec:form_decp}:
\begin{equation}
\hat{\tau}_{S_1} = \tau_0 + \hat{\tau}_1, \qquad
s_{S_1}^2 = \frac{1}{\hat{\tau}_{S_1}}, \qquad
\mu_{S_1} = \frac{\tau_0 \mu_0 + \hat{\tau}_1 \hat{\mu}_1}{\hat{\tau}_{S_1}},
\end{equation}
where $\tau_0 = 1/s_0^2$ and $\hat{\tau}_1 = 1/\hat{s}_1^2$.

\subsubsection{Stage 2: Adolescents}

Construct the prior for the adolescent analysis as a robust mixture of the informative component from Stage~1 and the unit-information component, with borrowing weight $w_2=0.5$:
\begin{equation}
\pi(\mu_2 \mid \theta_1) = 0.5\cdot \mathcal{N}\big(\mu_{S_1}, s_{S_1}\big) + 0.5\cdot \mathcal{N}(0,\, 2).
\end{equation}
After observing the adolescent summary $\theta_2 = (\hat{\mu}_2, \hat{s}_2) = (-0.58,\, 0.32)$, the posterior is a two-component mixture on admissible adjacent-only paths $\mathcal{P}_{1:2} = \{S_{1:2}=\{1,2\},\; S_2=\{2\}\}$:
\begin{equation}
\pi(\mu_2 \mid \theta_{1:2}) = w_2^\star\, \mathcal{N}\big(\mu_{S_{1:2}}, s_{S_{1:2}}\big) + (1 - w_2^\star)\, \mathcal{N}\big(\mu_{S_2}, s_{S_2}\big),
\end{equation}
where $(\mu_{S_{1:2}}, s_{S_{1:2}})$ and $(\mu_{S_2}, s_{S_2})$ are obtained by precision aggregation for $S\in\{S_{1:2}, S_2\}$. The top-level mixture weight $w_2^\star$ is updated using predictive marginal likelihoods as specified in Section~\ref{sec:form_decp}. In this case, the close agreement between the log-hazard ratios of adults and adolescents ($-0.62$ and $-0.58$, respectively) produces strong support for the informative component, resulting in posterior path weights $\mathcal{W}_{S_{1:2}} = w_2^\star \approx 0.86$ and $\mathcal{W}_{S_2} = 1 - w_2^\star \approx 0.14$.

\subsubsection{Stage 3: Paediatrics}

Form the prior for the paediatric analysis by combining the informative component (itself a mixture over adjacent paths up to Stage~2) with the unit-information component, with borrowing weight $w_3=0.5$:
\begin{equation}
\pi(\mu_3 \mid \theta_{1:2}) = 0.5 \bigg[\mathcal{W}_{S_{1:2}}\, \mathcal{N}\big(\mu_{S_{1:2}}, s_{S_{1:2}}\big) + \mathcal{W}_{S_2}\, \mathcal{N}\big(\mu_{S_2}, s_{S_2}\big) \bigg] + 0.5\, \mathcal{N}(0,\, 2)
\end{equation}
with $\mathcal{W}_{S_{1:2}} \approx 0.86$ and $\mathcal{W}_{S_2} \approx 0.14$ as determined in Stage~2.

After observing the paediatric summary $\theta_3 = (\hat{\mu}_3, \hat{s}_3) = (-0.51,\, 0.207)$, the posterior becomes a three-component mixture over admissible adjacent-only paths $\mathcal{P}_{1:3} = \{S_{1:3}=\{1,2,3\},\; S_{2:3}=\{2,3\},\; S_3=\{3\}\}$:
\begin{equation}
\pi(\mu_3 \mid \theta_{1:3}) = w_3^\star \Big[ w_2^\star\, \mathcal{N}\big(\mu_{S_{1:3}}, s_{S_{1:3}}\big) + (1-w_2^\star)\, \mathcal{N}\big(\mu_{S_{2:3}}, s_{S_{2:3}}\big) \Big] + (1 - w_3^\star)\, \mathcal{N}\big(\mu_{S_3}, s_{S_3}\big),
\end{equation}
where $(\mu_{S_{1:3}}, s_{S_{1:3}})$, $(\mu_{S_{2:3}}, s_{S_{2:3}})$ and $(\mu_{S_3}, s_{S_3})$ are obtained by precision aggregation for $S\in\{S_{1:3}, S_{2:3}, S_3\}$, and $w_3^\star$ is updated via predictive marginal likelihoods (Section~\ref{sec:form_decp}). The full posterior path weights are $\mathcal{W}_{S_{1:3}} = w_2^\star\, w_3^\star \approx 0.75$, $\mathcal{W}_{S_{2:3}} = (1-w_2^\star)\, w_3^\star \approx 0.13$, and $\mathcal{W}_{S_3} = 1 - w_3^\star \approx 0.12$.

Equivalently, the posterior can be written in expanded form:
\begin{equation}
\pi(\mu_3 \mid \theta_{1:3}) =
\underbrace{w_3^\star \mathcal{W}_{S_{1:3}}}_{\approx\, 0.75}\, \mathcal{N}\big(\mu_{S_{1:3}}, s_{S_{1:3}}\big) + \underbrace{w_3^\star \mathcal{W}_{S_{2:3}}}_{\approx\, 0.13}\, \mathcal{N}\big(\mu_{S_{2:3}}, s_{S_{2:3}}\big) + \underbrace{(1 - w_3^\star)}_{\approx\, 0.12}\, \mathcal{N}\big(\mu_{S_3}, s_{S_3}\big)
\end{equation}

The resulting posterior summaries for all three stages are given in Table~\ref{tab:case_study_res}.

\begin{table}[htb] 
\centering \begin{tabular}{|c|c|c|c|} 
\hline 
Population & Posterior Median HR (95\% CrI) & Posterior ESS & ESS \%\\ 
\hline 
Adult (18-66 yr) & 0.54 (0.40, 0.74) & 163 & 0.6\% \\ 
Adolescents (11-17 yr) & 0.55 (0.39, 0.79) & 154 & 74.3\%\\ 
Children (5-10 yr) & 0.57 (0.44, 0.77) & 219 & 57.3\%\\ 
\hline
\end{tabular} 
\caption{\emph{Posterior summaries under the RBSB approach.}} 
\label{tab:case_study_res} 
\end{table}

\subsection{Interpretation of the posterior mixture}

The posterior distribution for the log-hazard ratio in the paediatric population is a weighted combination of the admissible adjacent-only paths through the accumulated evidence:

\begin{enumerate}
\item Path $S_{1:3}$: borrowing from all three age strata (Adults $\to$ Adolescents $\to$ Paediatrics), with weight $w_3^\star w_2^\star \approx 0.75$,
\item Path $S_{2:3}$: borrowing from Adolescents and Paediatrics only, with weight $w_3^\star (1 - w_2^\star) \approx 0.13$,
\item Path $S_{3}$: using only the Paediatric data, with weight $1 - w_3^\star \approx 0.12$.
\end{enumerate}

The high weight assigned to the full path $S_{1:3}$ reflects the consistent treatment effects observed in all three age strata (log-HRs of $-0.62$, $-0.58$, and $-0.51$). Each component reflects a plausible pathway for information flow, either from all three strata (path $S_{1:3}$), from Adolescents and Paediatrics (path $S_{2:3}$), or from Paediatrics alone (path $S_3$), consistent with the adjacent-only borrowing scheme. The contribution of longer paths decays multiplicatively through the updated weights $w_2^\star$ and $w_3^\star$, attenuating borrowing when discordant evidence arises \cite{Schmidli,GiveDataAChance,Viele}. In particular, there is no mixture component that pools only Adults and Paediatrics (bypassing Adolescents); this enforces the adjacency constraint and mirrors the clinical intuition that Adolescents act as a bridge between Adults and Paediatrics.

As shown in Table~\ref{tab:case_study_res}, posterior median hazard ratios are below~1 in all strata, with the narrowest credible interval in adults (0.40--0.74), consistent with that subgroup's larger sample. RBSB tightens credible intervals in adolescents and children relative to standalone analyses: the adolescent posterior ESS of~154 (ESS\% = 74.3\%) indicates that borrowing from adults effectively augments the 40-event adolescent subgroup, while the paediatric ESS of~219 (ESS\% = 57.3\%) reflects borrowing moderated by the slightly attenuated log-hazard ratio in children. Adult ESS\% is 0.6\%, confirming that the first stage operates essentially standalone under the unit-information prior, as intended.

\section{Discussion}
The emphasis in this paper on adjacent borrowing for paediatric and adolescent programmes meets a genuine need and is consistent with regulatory concerns about unrealistic extrapolation in widely differing populations. By constraining information sharing to proceed only through neighbouring links (for example, adult $\rightarrow$ adolescent $\rightarrow$ paediatric), the proposed Robust Bayesian Sequential Borrowing (RBSB) framework puts this principle into practice and provides a clear, defensible method to temper prior influence when exchangeability is in doubt.

The key conceptual innovation is the path‑dependent construction that only allows borrowing between adjacent studies, setting this method apart from symmetric pooling strategies such as MAP and standard hierarchical random‑effects models. By explicitly encoding both the programme order and the anticipated biological proximity, the approach prevents information from “jumping” in non-adjacent studies, thus limiting unrealistic extrapolation and diminishing prior-data conflict. Its robust stepwise mixture formulation, with mixture weights updated by the data, enables borrowing to diminish automatically when new data are discordant, yet it still draws effectively on consistent adjacent information when it is compatible. From a decision‑making point of view, the framework aligns naturally with sequential Go/No‑Go rules and supports the assessment of Type I error, power and assurance at both the individual study and the overall level of the programme. This can improve study assurance and optimise resource use, especially in settings with limited power. Path‑dependent borrowing also helps explain why operating characteristics may vary across stages; the model explicitly represents these differences and allows them to be adjusted in advance via specified weights and unit‑information scales.

The proposed approach has many possible applications to synthesise information from sources with heterogeneous exchangeability, including age‑stratified trials, bridging between development phases, incorporating control data from registries, extrapolating effects between drug classes and regional programmes.

Our case study used concurrent data from the START trial, treating the age strata as if they had been studied sequentially. In true sequential programmes, RBSB would allow adaptive decision-making at each stage, potentially improving efficiency and reducing sample size requirements in later populations.


A complication of conditional metrics is what van Zwet et al. refer to as the winner’s curse\cite{winners_curse2021}. When results are filtered by statistical significance, as is implicit when conditioning on prior‑stage success, estimates in the selected set tend to be inflated, and nominal intervals can under‑cover the true effect. This arises from selection bias, where conditioning on significance preferentially retains larger observed effects even when the underlying signal is modest. van Zwet et al. recommend applying shrinkage to counter this bias, improving the calibration of effect estimates and interval coverage \cite{winners_curse2021}. Although our framework is Bayesian, it incorporates dynamic shrinkage through robust mixture priors that moderate the influence of prior evidence when discordance is detected. In sequential applications, this data-driven attenuation reduces the impact of the winner's curse on conditional summaries and supports more reliable decision making.

The Normal prior $\mathcal{N}(\hat{\mu}, \hat{s})$ is one possible choice at each time point. More flexible prior distributions, such as those arising from meta-analytic methods described by Schmidli et al. \cite{Schmidli}, can also be used. In this case, the prior at time $i$ can be specified as a mixture of Normal distributions, allowing the synthesis of evidence from multiple studies into a meta‑analytic predictive prior (MAP). For example, multiple adult trials can be combined to form a meta‑analytic prior that can then be extrapolated to neighbouring subpopulations. This makes it possible to adapt the borrowing structure at each time point to best match the available data and the desired level of generalisability, whether that is combining single‑study information or synthesising broader meta‑analytic evidence.

Assurance provides a prospective measure of success under a design prior and can be used to compare borrowing strategies and inform Go/No‑Go thresholds throughout the programme. In practice, the assurance and expected operating characteristics can be computed by simulation to calibrate weights, unit‑information scales, and decision rules to pre‑specified programme‑level criteria.

A distinction is warranted between conditional metrics, which are conditional on prior successes, and joint metrics, which require simultaneous success across stages. Conditional power at later stages can appear high when earlier stages have succeeded, yet the overall probability of multi-stage success declines multiplicatively with each additional requirement. In practice, “pushing problems downstream" reduces assurance at the programme level, because even modest stage-wise shortfalls compound when considered jointly. This is evident in our simulations, where designs with strong borrowing show high conditional power but materially lower joint success once sequential dependencies are accounted for. Accordingly, planning and business decisions should be based on joint (or programme-level) operating characteristics rather than conditional summaries alone, with pre-specified Go/No-Go rules calibrated to achieve acceptable joint power and to control joint false positives across the intended sequence. Where appropriate, resource allocation and sample size should be optimised at earlier stages to avoid loading risk into later, smaller studies, because early gains (or losses) propagate through the programme.

Future research could investigate interim analysis within RBSB, adaptive sample sizes for paediatric studies designed while adolescent studies are ongoing, and optimisation of resources across trials (for example, allocation of sample size to maximise programme‑level assurance subject to Type I error constraints, or using a utility maximisation objective). Extensions to wider evidence networks, such as multi‑source MAP priors with adjacency constraints, would further broaden applicability while maintaining disciplined borrowing.


\section*{Data Availability}

The simulation code and results are available at \url{https://github.com/AstraZeneca/RBSB}.

\appendix

\section{Proof of weights summing to 1}\label{app.proof}

Proof that \[
\sum_{S \in \mathcal{P}_{1:j}} \mathcal{W}_{S} = 1.
\]

Let $w^\star_\ell \in [0,1]$ for $\ell=1,\dots,j$ denote the posterior mixture weights at each step.
For contiguous paths $S_{i:j}$ with $1 \le i < j$, define
\[
W_{S_{i:j}} = (1 - w^\star_i)\prod_{\ell=i+1}^{j} w^\star_\ell,
\qquad
W_{\{j\}} = (1 - w^\star_j).
\]
Then
\[
\sum_{S \in \mathcal{P}_{1:j}} W_S
= (1 - w^\star_j)
+ \sum_{i=1}^{j-1} (1 - w^\star_i)\prod_{\ell=i+1}^{j} w^\star_\ell
= 1,
\]
which follows by the telescoping identity,
$1 = (1 - w^\star_j) + w^\star_j(1 - w^\star_{j-1}) + \cdots + w^\star_j w^\star_{j-1}\cdots w^\star_2(1 - w^\star_1)$
or by induction on $j$.

\section{Assurance}\label{app.assurance}

\subsection{Marginal Assurance}

This prior can be seen as given by nested robust mixture priors. This is also a coherent prior per de Finetti's criteria \cite{de1974theory}. The assurance is then 

\begin{equation} \label{eq:marginal_assurnace}
\begin{aligned}
\Omega_{\mathcal{P}_{i:j}} &=
 \Pr\left(D_j =1 ~\middle|\theta_{1:(j-1)},\,\mu_j \sim \pi(\mu_j\mid\theta_{1:(j-1)})\right) \\
 &=
  \int \Pr\left(D_j =1 ~\middle|\theta_{1:(j-1)},\,\mu_j\right)\pi(\mu_j \mid \theta_{1:(j-1)})\, d\mu_j
\end{aligned}
\end{equation}

\subsection{Conditional Assurance}

The assurance is then the conditional probability that we declare a study success given that the treatment effect comes from the prior distribution and that we have declared success at all previous points.

\begin{equation} \label{eq:conditional_assurnace}
\begin{aligned}
\Omega_{\mathcal{P}_{i:j}}^{(C)} 
 &=
  \int \Pr\left(D_j =1~\middle|~\bigcap_{l=i}^{j-1} D_l =1,\,\theta_{1:(j-1)},\,\mu_j\right)\pi(\mu_j \mid \theta_{1:(j-1)})\, d\mu_j
\end{aligned}
\end{equation}

\subsection{Joint Assurance}
Joint assurance for the entire sequence is defined as

\begin{equation} \label{eq:joint_assurnace}
\begin{aligned}
\Omega_{\mathcal{P}_{i:j}}^{(J)}
 &=
  \int \Pr\left(\bigcap_{l=i}^{j} D_l =1 ~\middle|\,\theta_{1:(j-1)},\,\mu_j\right)\pi(\mu_j \mid \theta_{1:(j-1)})\, d\mu_j
\end{aligned}
\end{equation}

\section{Simulation Results} \label{app.sim_result}

\input{tables/marginal_rr}

\input{tables/conditional_rr}

\input{tables/joint_rr}

\input{tables/ess}

\input{tables/bias}

\clearpage
\newpage
\bibliographystyle{abbrv}
\bibliography{refs}

\end{document}

%% file: tables/marginal_rr.tex
\begin{table}[ht]
\centering
\begin{tabular}{lrrrrrc}
  \toprule
& \multicolumn{5}{c}{\textbf{Marginal Rej.\ Rate (\%)}} & \\ \cmidrule(lr){2-6}\textbf{Method} & \textbf{$k$=1} & \textbf{$k$=2} & \textbf{$k$=3} & \textbf{$k$=4} & \textbf{$k$=5} & \textbf{Assump} \\ 
  \midrule
\addlinespace[2pt]\multicolumn{7}{l}{\textbf{Setting 1}}\\Standalone & 2.6 & 2.4 & 2.4 & 2.4 & 2.4 & 0, 0, 0, 0, 0 \\ 
  Full pooling Bayesian borrowing & 2.6 & 1.7 & 1.5 & 1.3 & 1.2 & 0, 0, 0, 0, 0 \\ 
  Adjacent Bayesian borrowing & 2.6 & 1.7 & 2.1 & 1.9 & 2.1 & 0, 0, 0, 0, 0 \\ 
  RBSB & 2.6 & 1.7 & 1.2 & 0.9 & 0.9 & 0, 0, 0, 0, 0 \\ 
   \addlinespace[2pt]\multicolumn{7}{l}{\textbf{Setting 2}}\\Standalone & 87.3 & 2.4 & 2.4 & 2.4 & 2.4 & $\delta$, 0, 0, 0, 0 \\ 
  Full pooling Bayesian borrowing & 87.6 & 12.9 & 13.4 & 12.5 & 11.3 & $\delta$, 0, 0, 0, 0 \\ 
  Adjacent Bayesian borrowing & 87.6 & 12.9 & 2.1 & 1.9 & 2.1 & $\delta$, 0, 0, 0, 0 \\ 
  RBSB & 87.6 & 13.1 & 4.4 & 2.1 & 1.6 & $\delta$, 0, 0, 0, 0 \\ 
   \addlinespace[2pt]\multicolumn{7}{l}{\textbf{Setting 3}}\\Standalone & 87.3 & 47.2 & 2.4 & 2.4 & 2.4 & $\delta$, $\delta$, 0, 0, 0 \\ 
  Full pooling Bayesian borrowing & 87.6 & 76.1 & 14.1 & 14.8 & 15.3 & $\delta$, $\delta$, 0, 0, 0 \\ 
  Adjacent Bayesian borrowing & 87.6 & 76.1 & 7.4 & 1.9 & 2.1 & $\delta$, $\delta$, 0, 0, 0 \\ 
  RBSB & 87.6 & 76.1 & 11.5 & 3.8 & 2.2 & $\delta$, $\delta$, 0, 0, 0 \\ 
   \addlinespace[2pt]\multicolumn{7}{l}{\textbf{Setting 4}}\\Standalone & 87.3 & 47.2 & 47.2 & 2.4 & 2.4 & $\delta$, $\delta$, $\delta$, 0, 0 \\ 
  Full pooling Bayesian borrowing & 87.6 & 76.1 & 78.8 & 14.2 & 16.4 & $\delta$, $\delta$, $\delta$, 0, 0 \\ 
  Adjacent Bayesian borrowing & 87.6 & 76.1 & 64.1 & 7.0 & 2.1 & $\delta$, $\delta$, $\delta$, 0, 0 \\ 
  RBSB & 87.6 & 76.1 & 73.8 & 10.8 & 4.2 & $\delta$, $\delta$, $\delta$, 0, 0 \\ 
   \addlinespace[2pt]\multicolumn{7}{l}{\textbf{Setting 5}}\\Standalone & 87.3 & 47.2 & 47.2 & 47.4 & 2.4 & $\delta$, $\delta$, $\delta$, $\delta$, 0 \\ 
  Full pooling Bayesian borrowing & 87.6 & 76.1 & 78.8 & 80.4 & 15.1 & $\delta$, $\delta$, $\delta$, $\delta$, 0 \\ 
  Adjacent Bayesian borrowing & 87.6 & 76.1 & 64.1 & 64.9 & 7.4 & $\delta$, $\delta$, $\delta$, $\delta$, 0 \\ 
  RBSB & 87.6 & 76.1 & 73.8 & 72.9 & 11.0 & $\delta$, $\delta$, $\delta$, $\delta$, 0 \\ 
   \addlinespace[2pt]\multicolumn{7}{l}{\textbf{Setting 6}}\\Standalone & 87.3 & 47.2 & 47.2 & 47.4 & 46.9 & $\delta$, $\delta$, $\delta$, $\delta$, $\delta$ \\ 
  Full pooling Bayesian borrowing & 87.6 & 76.1 & 78.8 & 80.4 & 80.8 & $\delta$, $\delta$, $\delta$, $\delta$, $\delta$ \\ 
  Adjacent Bayesian borrowing & 87.6 & 76.1 & 64.1 & 64.9 & 64.5 & $\delta$, $\delta$, $\delta$, $\delta$, $\delta$ \\ 
  RBSB & 87.6 & 76.1 & 73.8 & 72.9 & 73.0 & $\delta$, $\delta$, $\delta$, $\delta$, $\delta$ \\ 
   \addlinespace[2pt]\multicolumn{7}{l}{\textbf{Setting 7}}\\Standalone & 87.3 & 47.2 & 20.3 & 2.4 & 2.5 & $\delta$, $\delta$, $0.6\!\cdot\!\delta$, 0, 0 \\ 
  Full pooling Bayesian borrowing & 87.6 & 76.1 & 52.6 & 14.8 & 16.3 & $\delta$, $\delta$, $0.6\!\cdot\!\delta$, 0, 0 \\ 
  Adjacent Bayesian borrowing & 87.6 & 76.1 & 36.3 & 5.1 & 2.0 & $\delta$, $\delta$, $0.6\!\cdot\!\delta$, 0, 0 \\ 
  RBSB & 87.6 & 76.1 & 46.3 & 8.5 & 3.6 & $\delta$, $\delta$, $0.6\!\cdot\!\delta$, 0, 0 \\ 
   \addlinespace[2pt]\multicolumn{7}{l}{\textbf{Setting 8}}\\Standalone & 87.3 & 47.2 & 20.3 & 20.0 & 2.5 & $\delta$, $\delta$, $0.6\!\cdot\!\delta$, $0.6\!\cdot\!\delta$, 0 \\ 
  Full pooling Bayesian borrowing & 87.6 & 76.1 & 52.6 & 55.0 & 16.6 & $\delta$, $\delta$, $0.6\!\cdot\!\delta$, $0.6\!\cdot\!\delta$, 0 \\ 
  Adjacent Bayesian borrowing & 87.6 & 76.1 & 36.3 & 28.2 & 5.5 & $\delta$, $\delta$, $0.6\!\cdot\!\delta$, $0.6\!\cdot\!\delta$, 0 \\ 
  RBSB & 87.6 & 76.1 & 46.3 & 38.9 & 8.4 & $\delta$, $\delta$, $0.6\!\cdot\!\delta$, $0.6\!\cdot\!\delta$, 0 \\ 
   \addlinespace[2pt]\multicolumn{7}{l}{\textbf{Setting 9}}\\Standalone & 87.3 & 47.2 & 20.3 & 20.0 & 18.4 & $\delta$, $\delta$, $0.6\!\cdot\!\delta$, $0.6\!\cdot\!\delta$, $0.6\!\cdot\!\delta$ \\ 
  Full pooling Bayesian borrowing & 87.6 & 76.1 & 52.6 & 55.0 & 52.3 & $\delta$, $\delta$, $0.6\!\cdot\!\delta$, $0.6\!\cdot\!\delta$, $0.6\!\cdot\!\delta$ \\ 
  Adjacent Bayesian borrowing & 87.6 & 76.1 & 36.3 & 28.2 & 25.8 & $\delta$, $\delta$, $0.6\!\cdot\!\delta$, $0.6\!\cdot\!\delta$, $0.6\!\cdot\!\delta$ \\ 
  RBSB & 87.6 & 76.1 & 46.3 & 38.9 & 33.2 & $\delta$, $\delta$, $0.6\!\cdot\!\delta$, $0.6\!\cdot\!\delta$, $0.6\!\cdot\!\delta$ \\ 
   \bottomrule
\end{tabular}
\caption{Marginal rejection rates for different settings (10\,000 simulations).} 
\label{tab:marginal}
\end{table}

%% file: tables/conditional_rr.tex
\begin{table}[ht]
\centering
\begin{tabular}{lrrrrc}
  \toprule
& \multicolumn{4}{c}{\textbf{Cond.\ Rej.\ Rate (\%)}} & \\ \cmidrule(lr){2-5}\textbf{Method} & \textbf{$k$=2} & \textbf{$k$=3} & \textbf{$k$=4} & \textbf{$k$=5} & \textbf{Assump} \\ 
  \midrule
\addlinespace[2pt]\multicolumn{6}{l}{\textbf{Setting 1}}\\Standalone & 3.1 & 12.5 & 0.0 & NA & 0, 0, 0, 0, 0 \\ 
  Full pooling Bayesian borrowing & 17.4 & 23.9 & 0.0 & NA & 0, 0, 0, 0, 0 \\ 
  Adjacent Bayesian borrowing & 17.4 & 15.2 & 0.0 & NA & 0, 0, 0, 0, 0 \\ 
  RBSB & 17.4 & 19.6 & 0.0 & NA & 0, 0, 0, 0, 0 \\ 
   \addlinespace[2pt]\multicolumn{6}{l}{\textbf{Setting 2}}\\Standalone & 2.5 & 2.8 & 0.0 & NA & $\delta$, 0, 0, 0, 0 \\ 
  Full pooling Bayesian borrowing & 13.8 & 15.1 & 17.0 & 16.1 & $\delta$, 0, 0, 0, 0 \\ 
  Adjacent Bayesian borrowing & 13.8 & 7.5 & 8.9 & 12.5 & $\delta$, 0, 0, 0, 0 \\ 
  RBSB & 14.0 & 13.1 & 16.1 & 15.4 & $\delta$, 0, 0, 0, 0 \\ 
   \addlinespace[2pt]\multicolumn{6}{l}{\textbf{Setting 3}}\\Standalone & 47.4 & 2.4 & 3.0 & 0.0 & $\delta$, $\delta$, 0, 0, 0 \\ 
  Full pooling Bayesian borrowing & 78.0 & 13.8 & 13.6 & 20.3 & $\delta$, $\delta$, 0, 0, 0 \\ 
  Adjacent Bayesian borrowing & 78.0 & 8.6 & 7.8 & 8.7 & $\delta$, $\delta$, 0, 0, 0 \\ 
  RBSB & 78.0 & 12.7 & 12.5 & 15.7 & $\delta$, $\delta$, 0, 0, 0 \\ 
   \addlinespace[2pt]\multicolumn{6}{l}{\textbf{Setting 4}}\\Standalone & 47.4 & 47.3 & 2.0 & 0.0 & $\delta$, $\delta$, $\delta$, 0, 0 \\ 
  Full pooling Bayesian borrowing & 78.0 & 78.8 & 13.4 & 15.0 & $\delta$, $\delta$, $\delta$, 0, 0 \\ 
  Adjacent Bayesian borrowing & 78.0 & 69.7 & 8.9 & 8.2 & $\delta$, $\delta$, $\delta$, 0, 0 \\ 
  RBSB & 78.0 & 77.4 & 12.0 & 13.8 & $\delta$, $\delta$, $\delta$, 0, 0 \\ 
   \addlinespace[2pt]\multicolumn{6}{l}{\textbf{Setting 5}}\\Standalone & 47.4 & 47.3 & 47.2 & 2.5 & $\delta$, $\delta$, $\delta$, $\delta$, 0 \\ 
  Full pooling Bayesian borrowing & 78.0 & 78.8 & 80.1 & 13.5 & $\delta$, $\delta$, $\delta$, $\delta$, 0 \\ 
  Adjacent Bayesian borrowing & 78.0 & 69.7 & 72.3 & 9.0 & $\delta$, $\delta$, $\delta$, $\delta$, 0 \\ 
  RBSB & 78.0 & 77.4 & 78.0 & 12.2 & $\delta$, $\delta$, $\delta$, $\delta$, 0 \\ 
   \addlinespace[2pt]\multicolumn{6}{l}{\textbf{Setting 6}}\\Standalone & 47.4 & 47.3 & 47.2 & 48.8 & $\delta$, $\delta$, $\delta$, $\delta$, $\delta$ \\ 
  Full pooling Bayesian borrowing & 78.0 & 78.8 & 80.1 & 79.2 & $\delta$, $\delta$, $\delta$, $\delta$, $\delta$ \\ 
  Adjacent Bayesian borrowing & 78.0 & 69.7 & 72.3 & 71.3 & $\delta$, $\delta$, $\delta$, $\delta$, $\delta$ \\ 
  RBSB & 78.0 & 77.4 & 78.0 & 77.3 & $\delta$, $\delta$, $\delta$, $\delta$, $\delta$ \\ 
   \addlinespace[2pt]\multicolumn{6}{l}{\textbf{Setting 7}}\\Standalone & 47.4 & 20.9 & 2.1 & 0.0 & $\delta$, $\delta$, $0.6\!\cdot\!\delta$, 0, 0 \\ 
  Full pooling Bayesian borrowing & 78.0 & 52.5 & 13.6 & 15.4 & $\delta$, $\delta$, $0.6\!\cdot\!\delta$, 0, 0 \\ 
  Adjacent Bayesian borrowing & 78.0 & 40.6 & 8.8 & 5.8 & $\delta$, $\delta$, $0.6\!\cdot\!\delta$, 0, 0 \\ 
  RBSB & 78.0 & 50.0 & 12.0 & 13.2 & $\delta$, $\delta$, $0.6\!\cdot\!\delta$, 0, 0 \\ 
   \addlinespace[2pt]\multicolumn{6}{l}{\textbf{Setting 8}}\\Standalone & 47.4 & 20.9 & 19.2 & 0.6 & $\delta$, $\delta$, $0.6\!\cdot\!\delta$, $0.6\!\cdot\!\delta$, 0 \\ 
  Full pooling Bayesian borrowing & 78.0 & 52.5 & 53.6 & 15.2 & $\delta$, $\delta$, $0.6\!\cdot\!\delta$, $0.6\!\cdot\!\delta$, 0 \\ 
  Adjacent Bayesian borrowing & 78.0 & 40.6 & 42.1 & 8.4 & $\delta$, $\delta$, $0.6\!\cdot\!\delta$, $0.6\!\cdot\!\delta$, 0 \\ 
  RBSB & 78.0 & 50.0 & 49.8 & 13.9 & $\delta$, $\delta$, $0.6\!\cdot\!\delta$, $0.6\!\cdot\!\delta$, 0 \\ 
   \addlinespace[2pt]\multicolumn{6}{l}{\textbf{Setting 9}}\\Standalone & 47.4 & 20.9 & 19.2 & 10.8 & $\delta$, $\delta$, $0.6\!\cdot\!\delta$, $0.6\!\cdot\!\delta$, $0.6\!\cdot\!\delta$ \\ 
  Full pooling Bayesian borrowing & 78.0 & 52.5 & 53.6 & 50.6 & $\delta$, $\delta$, $0.6\!\cdot\!\delta$, $0.6\!\cdot\!\delta$, $0.6\!\cdot\!\delta$ \\ 
  Adjacent Bayesian borrowing & 78.0 & 40.6 & 42.1 & 38.8 & $\delta$, $\delta$, $0.6\!\cdot\!\delta$, $0.6\!\cdot\!\delta$, $0.6\!\cdot\!\delta$ \\ 
  RBSB & 78.0 & 50.0 & 49.8 & 47.6 & $\delta$, $\delta$, $0.6\!\cdot\!\delta$, $0.6\!\cdot\!\delta$, $0.6\!\cdot\!\delta$ \\ 
   \bottomrule
\end{tabular}
\caption{Conditional rejection rates for different settings (10\,000 simulations).} 
\label{tab:cond}
\end{table}

%% file: tables/joint_rr.tex
\begin{table}[ht]
\centering
\begin{tabular}{lrrrrrc}
  \toprule
& \multicolumn{5}{c}{\textbf{Joint Rej.\ Rate (\%)}} & \\ \cmidrule(lr){2-6}\textbf{Method} & \textbf{$k$=1} & \textbf{$k$=2} & \textbf{$k$=3} & \textbf{$k$=4} & \textbf{$k$=5} & \textbf{Assump} \\ 
  \midrule
\addlinespace[2pt]\multicolumn{7}{l}{\textbf{Setting 1}}\\Standalone & 2.6 & 0.1 & 0.0 & 0.0 & 0.0 & 0, 0, 0, 0, 0 \\ 
  Full pooling Bayesian borrowing & 2.6 & 0.5 & 0.1 & 0.0 & 0.0 & 0, 0, 0, 0, 0 \\ 
  Adjacent Bayesian borrowing & 2.6 & 0.5 & 0.1 & 0.0 & 0.0 & 0, 0, 0, 0, 0 \\ 
  RBSB & 2.6 & 0.5 & 0.1 & 0.0 & 0.0 & 0, 0, 0, 0, 0 \\ 
   \addlinespace[2pt]\multicolumn{7}{l}{\textbf{Setting 2}}\\Standalone & 87.3 & 2.2 & 0.1 & 0.0 & 0.0 & $\delta$, 0, 0, 0, 0 \\ 
  Full pooling Bayesian borrowing & 87.6 & 12.1 & 1.8 & 0.3 & 0.1 & $\delta$, 0, 0, 0, 0 \\ 
  Adjacent Bayesian borrowing & 87.6 & 12.1 & 0.9 & 0.1 & 0.0 & $\delta$, 0, 0, 0, 0 \\ 
  RBSB & 87.6 & 12.3 & 1.6 & 0.3 & 0.0 & $\delta$, 0, 0, 0, 0 \\ 
   \addlinespace[2pt]\multicolumn{7}{l}{\textbf{Setting 3}}\\Standalone & 87.3 & 41.4 & 1.0 & 0.0 & 0.0 & $\delta$, $\delta$, 0, 0, 0 \\ 
  Full pooling Bayesian borrowing & 87.6 & 68.3 & 9.4 & 1.3 & 0.3 & $\delta$, $\delta$, 0, 0, 0 \\ 
  Adjacent Bayesian borrowing & 87.6 & 68.3 & 5.9 & 0.5 & 0.0 & $\delta$, $\delta$, 0, 0, 0 \\ 
  RBSB & 87.6 & 68.3 & 8.7 & 1.1 & 0.2 & $\delta$, $\delta$, 0, 0, 0 \\ 
   \addlinespace[2pt]\multicolumn{7}{l}{\textbf{Setting 4}}\\Standalone & 87.3 & 41.4 & 19.6 & 0.4 & 0.0 & $\delta$, $\delta$, $\delta$, 0, 0 \\ 
  Full pooling Bayesian borrowing & 87.6 & 68.3 & 53.9 & 7.2 & 1.1 & $\delta$, $\delta$, $\delta$, 0, 0 \\ 
  Adjacent Bayesian borrowing & 87.6 & 68.3 & 47.7 & 4.3 & 0.4 & $\delta$, $\delta$, $\delta$, 0, 0 \\ 
  RBSB & 87.6 & 68.3 & 52.9 & 6.4 & 0.9 & $\delta$, $\delta$, $\delta$, 0, 0 \\ 
   \addlinespace[2pt]\multicolumn{7}{l}{\textbf{Setting 5}}\\Standalone & 87.3 & 41.4 & 19.6 & 9.2 & 0.2 & $\delta$, $\delta$, $\delta$, $\delta$, 0 \\ 
  Full pooling Bayesian borrowing & 87.6 & 68.3 & 53.9 & 43.1 & 5.8 & $\delta$, $\delta$, $\delta$, $\delta$, 0 \\ 
  Adjacent Bayesian borrowing & 87.6 & 68.3 & 47.7 & 34.5 & 3.1 & $\delta$, $\delta$, $\delta$, $\delta$, 0 \\ 
  RBSB & 87.6 & 68.3 & 52.9 & 41.2 & 5.1 & $\delta$, $\delta$, $\delta$, $\delta$, 0 \\ 
   \addlinespace[2pt]\multicolumn{7}{l}{\textbf{Setting 6}}\\Standalone & 87.3 & 41.4 & 19.6 & 9.2 & 4.5 & $\delta$, $\delta$, $\delta$, $\delta$, $\delta$ \\ 
  Full pooling Bayesian borrowing & 87.6 & 68.3 & 53.9 & 43.1 & 34.2 & $\delta$, $\delta$, $\delta$, $\delta$, $\delta$ \\ 
  Adjacent Bayesian borrowing & 87.6 & 68.3 & 47.7 & 34.5 & 24.6 & $\delta$, $\delta$, $\delta$, $\delta$, $\delta$ \\ 
  RBSB & 87.6 & 68.3 & 52.9 & 41.2 & 31.9 & $\delta$, $\delta$, $\delta$, $\delta$, $\delta$ \\ 
   \addlinespace[2pt]\multicolumn{7}{l}{\textbf{Setting 7}}\\Standalone & 87.3 & 41.4 & 8.6 & 0.2 & 0.0 & $\delta$, $\delta$, $0.6\!\cdot\!\delta$, 0, 0 \\ 
  Full pooling Bayesian borrowing & 87.6 & 68.3 & 35.9 & 4.9 & 0.8 & $\delta$, $\delta$, $0.6\!\cdot\!\delta$, 0, 0 \\ 
  Adjacent Bayesian borrowing & 87.6 & 68.3 & 27.8 & 2.4 & 0.1 & $\delta$, $\delta$, $0.6\!\cdot\!\delta$, 0, 0 \\ 
  RBSB & 87.6 & 68.3 & 34.2 & 4.1 & 0.5 & $\delta$, $\delta$, $0.6\!\cdot\!\delta$, 0, 0 \\ 
   \addlinespace[2pt]\multicolumn{7}{l}{\textbf{Setting 8}}\\Standalone & 87.3 & 41.4 & 8.6 & 1.7 & 0.0 & $\delta$, $\delta$, $0.6\!\cdot\!\delta$, $0.6\!\cdot\!\delta$, 0 \\ 
  Full pooling Bayesian borrowing & 87.6 & 68.3 & 35.9 & 19.2 & 2.9 & $\delta$, $\delta$, $0.6\!\cdot\!\delta$, $0.6\!\cdot\!\delta$, 0 \\ 
  Adjacent Bayesian borrowing & 87.6 & 68.3 & 27.8 & 11.7 & 1.0 & $\delta$, $\delta$, $0.6\!\cdot\!\delta$, $0.6\!\cdot\!\delta$, 0 \\ 
  RBSB & 87.6 & 68.3 & 34.2 & 17.0 & 2.4 & $\delta$, $\delta$, $0.6\!\cdot\!\delta$, $0.6\!\cdot\!\delta$, 0 \\ 
   \addlinespace[2pt]\multicolumn{7}{l}{\textbf{Setting 9}}\\Standalone & 87.3 & 41.4 & 8.6 & 1.7 & 0.2 & $\delta$, $\delta$, $0.6\!\cdot\!\delta$, $0.6\!\cdot\!\delta$, $0.6\!\cdot\!\delta$ \\ 
  Full pooling Bayesian borrowing & 87.6 & 68.3 & 35.9 & 19.2 & 9.7 & $\delta$, $\delta$, $0.6\!\cdot\!\delta$, $0.6\!\cdot\!\delta$, $0.6\!\cdot\!\delta$ \\ 
  Adjacent Bayesian borrowing & 87.6 & 68.3 & 27.8 & 11.7 & 4.5 & $\delta$, $\delta$, $0.6\!\cdot\!\delta$, $0.6\!\cdot\!\delta$, $0.6\!\cdot\!\delta$ \\ 
  RBSB & 87.6 & 68.3 & 34.2 & 17.0 & 8.1 & $\delta$, $\delta$, $0.6\!\cdot\!\delta$, $0.6\!\cdot\!\delta$, $0.6\!\cdot\!\delta$ \\ 
   \bottomrule
\end{tabular}
\caption{Joint rejection rates for different settings (10\,000 simulations).} 
\label{tab:joint}
\end{table}

%% file: tables/ess.tex
\begin{table}[ht]
\centering
\begin{tabular}{lrrrrrc}
  \toprule
& \multicolumn{5}{c}{\textbf{ESS (\%)}} & \\ \cmidrule(lr){2-6}\textbf{Method} & \textbf{$k$=1} & \textbf{$k$=2} & \textbf{$k$=3} & \textbf{$k$=4} & \textbf{$k$=5} & \textbf{Assump} \\ 
  \midrule
\addlinespace[2pt]\multicolumn{7}{l}{\textbf{Setting 1}}\\Standalone & 0.00 & 0.00 & 0.00 & 0.00 & 0.00 & 0, 0, 0, 0, 0 \\ 
  Full pooling Bayesian borrowing & 0.62 & 58.23 & 66.39 & 72.21 & 76.06 & 0, 0, 0, 0, 0 \\ 
  Adjacent Bayesian borrowing & 0.62 & 58.23 & 32.85 & 33.20 & 33.23 & 0, 0, 0, 0, 0 \\ 
  RBSB & 0.62 & 58.75 & 57.86 & 57.92 & 57.92 & 0, 0, 0, 0, 0 \\ 
   \addlinespace[2pt]\multicolumn{7}{l}{\textbf{Setting 2}}\\Standalone & 0.00 & 0.00 & 0.00 & 0.00 & 0.00 & $\delta$, 0, 0, 0, 0 \\ 
  Full pooling Bayesian borrowing & 0.62 & 35.20 & 53.17 & 64.32 & 70.97 & $\delta$, 0, 0, 0, 0 \\ 
  Adjacent Bayesian borrowing & 0.62 & 35.20 & 32.85 & 33.20 & 33.23 & $\delta$, 0, 0, 0, 0 \\ 
  RBSB & 0.62 & 35.62 & 39.17 & 42.66 & 45.58 & $\delta$, 0, 0, 0, 0 \\ 
   \addlinespace[2pt]\multicolumn{7}{l}{\textbf{Setting 3}}\\Standalone & 0.00 & 0.00 & 0.00 & 0.00 & 0.00 & $\delta$, $\delta$, 0, 0, 0 \\ 
  Full pooling Bayesian borrowing & 0.62 & 58.55 & 43.40 & 58.01 & 66.83 & $\delta$, $\delta$, 0, 0, 0 \\ 
  Adjacent Bayesian borrowing & 0.62 & 58.55 & 15.78 & 33.20 & 33.23 & $\delta$, $\delta$, 0, 0, 0 \\ 
  RBSB & 0.62 & 59.07 & 37.37 & 39.47 & 41.96 & $\delta$, $\delta$, 0, 0, 0 \\ 
   \addlinespace[2pt]\multicolumn{7}{l}{\textbf{Setting 4}}\\Standalone & 0.00 & 0.00 & 0.00 & 0.00 & 0.00 & $\delta$, $\delta$, $\delta$, 0, 0 \\ 
  Full pooling Bayesian borrowing & 0.62 & 58.55 & 66.67 & 50.17 & 61.48 & $\delta$, $\delta$, $\delta$, 0, 0 \\ 
  Adjacent Bayesian borrowing & 0.62 & 58.55 & 33.12 & 16.05 & 33.23 & $\delta$, $\delta$, $\delta$, 0, 0 \\ 
  RBSB & 0.62 & 59.07 & 58.35 & 37.51 & 39.28 & $\delta$, $\delta$, $\delta$, 0, 0 \\ 
   \addlinespace[2pt]\multicolumn{7}{l}{\textbf{Setting 5}}\\Standalone & 0.00 & 0.00 & 0.00 & 0.00 & 0.00 & $\delta$, $\delta$, $\delta$, $\delta$, 0 \\ 
  Full pooling Bayesian borrowing & 0.62 & 58.55 & 66.67 & 72.46 & 55.04 & $\delta$, $\delta$, $\delta$, $\delta$, 0 \\ 
  Adjacent Bayesian borrowing & 0.62 & 58.55 & 33.12 & 33.48 & 16.03 & $\delta$, $\delta$, $\delta$, $\delta$, 0 \\ 
  RBSB & 0.62 & 59.07 & 58.35 & 58.51 & 37.87 & $\delta$, $\delta$, $\delta$, $\delta$, 0 \\ 
   \addlinespace[2pt]\multicolumn{7}{l}{\textbf{Setting 6}}\\Standalone & 0.00 & 0.00 & 0.00 & 0.00 & 0.00 & $\delta$, $\delta$, $\delta$, $\delta$, $\delta$ \\ 
  Full pooling Bayesian borrowing & 0.62 & 58.55 & 66.67 & 72.46 & 76.29 & $\delta$, $\delta$, $\delta$, $\delta$, $\delta$ \\ 
  Adjacent Bayesian borrowing & 0.62 & 58.55 & 33.12 & 33.48 & 33.51 & $\delta$, $\delta$, $\delta$, $\delta$, $\delta$ \\ 
  RBSB & 0.62 & 59.07 & 58.35 & 58.51 & 58.58 & $\delta$, $\delta$, $\delta$, $\delta$, $\delta$ \\ 
   \addlinespace[2pt]\multicolumn{7}{l}{\textbf{Setting 7}}\\Standalone & 0.00 & 0.00 & 0.00 & 0.00 & 0.00 & $\delta$, $\delta$, $0.6\!\cdot\!\delta$, 0, 0 \\ 
  Full pooling Bayesian borrowing & 0.62 & 58.55 & 62.56 & 53.62 & 68.93 & $\delta$, $\delta$, $0.6\!\cdot\!\delta$, 0, 0 \\ 
  Adjacent Bayesian borrowing & 0.62 & 58.55 & 29.74 & 26.32 & 36.69 & $\delta$, $\delta$, $0.6\!\cdot\!\delta$, 0, 0 \\ 
  RBSB & 0.62 & 59.07 & 54.63 & 42.03 & 47.47 & $\delta$, $\delta$, $0.6\!\cdot\!\delta$, 0, 0 \\ 
   \addlinespace[2pt]\multicolumn{7}{l}{\textbf{Setting 8}}\\Standalone & 0.00 & 0.00 & 0.00 & 0.00 & 0.00 & $\delta$, $\delta$, $0.6\!\cdot\!\delta$, $0.6\!\cdot\!\delta$, 0 \\ 
  Full pooling Bayesian borrowing & 0.62 & 58.55 & 62.56 & 70.03 & 66.19 & $\delta$, $\delta$, $0.6\!\cdot\!\delta$, $0.6\!\cdot\!\delta$, 0 \\ 
  Adjacent Bayesian borrowing & 0.62 & 58.55 & 29.74 & 33.30 & 30.43 & $\delta$, $\delta$, $0.6\!\cdot\!\delta$, $0.6\!\cdot\!\delta$, 0 \\ 
  RBSB & 0.62 & 59.07 & 54.63 & 54.28 & 50.02 & $\delta$, $\delta$, $0.6\!\cdot\!\delta$, $0.6\!\cdot\!\delta$, 0 \\ 
   \addlinespace[2pt]\multicolumn{7}{l}{\textbf{Setting 9}}\\Standalone & 0.00 & 0.00 & 0.00 & 0.00 & 0.00 & $\delta$, $\delta$, $0.6\!\cdot\!\delta$, $0.6\!\cdot\!\delta$, $0.6\!\cdot\!\delta$ \\ 
  Full pooling Bayesian borrowing & 0.62 & 58.55 & 62.56 & 70.03 & 77.65 & $\delta$, $\delta$, $0.6\!\cdot\!\delta$, $0.6\!\cdot\!\delta$, $0.6\!\cdot\!\delta$ \\ 
  Adjacent Bayesian borrowing & 0.62 & 58.55 & 29.74 & 33.30 & 36.80 & $\delta$, $\delta$, $0.6\!\cdot\!\delta$, $0.6\!\cdot\!\delta$, $0.6\!\cdot\!\delta$ \\ 
  RBSB & 0.62 & 59.07 & 54.63 & 54.28 & 58.71 & $\delta$, $\delta$, $0.6\!\cdot\!\delta$, $0.6\!\cdot\!\delta$, $0.6\!\cdot\!\delta$ \\ 
   \bottomrule
\end{tabular}
\caption{ESS percent for different settings (10\,000 simulations).} 
\label{tab:ess}
\end{table}

%% file: tables/bias.tex
\begin{table}[ht]
\centering
\begin{tabular}{lrrrrrc}
  \toprule
& \multicolumn{5}{c}{\textbf{Bias}} & \\ \cmidrule(lr){2-6}\textbf{Method} & \textbf{$k$=1} & \textbf{$k$=2} & \textbf{$k$=3} & \textbf{$k$=4} & \textbf{$k$=5} & \textbf{Assump} \\ 
  \midrule
\addlinespace[2pt]\multicolumn{7}{l}{\textbf{Setting 1}}\\Standalone & -0.00 & 0.00 & -0.00 & -0.00 & -0.00 & 0, 0, 0, 0, 0 \\ 
  Full pooling Bayesian borrowing & -0.00 & -0.00 & -0.00 & -0.00 & -0.00 & 0, 0, 0, 0, 0 \\ 
  Adjacent Bayesian borrowing & -0.00 & -0.00 & -0.00 & -0.00 & -0.00 & 0, 0, 0, 0, 0 \\ 
  RBSB & -0.00 & -0.00 & -0.00 & -0.00 & -0.00 & 0, 0, 0, 0, 0 \\ 
   \addlinespace[2pt]\multicolumn{7}{l}{\textbf{Setting 2}}\\Standalone & -0.00 & 0.00 & -0.00 & -0.00 & -0.00 & $\delta$, 0, 0, 0, 0 \\ 
  Full pooling Bayesian borrowing & -0.00 & 0.16 & 0.16 & 0.15 & 0.13 & $\delta$, 0, 0, 0, 0 \\ 
  Adjacent Bayesian borrowing & -0.00 & 0.16 & -0.00 & -0.00 & -0.00 & $\delta$, 0, 0, 0, 0 \\ 
  RBSB & -0.00 & 0.16 & 0.08 & 0.05 & 0.03 & $\delta$, 0, 0, 0, 0 \\ 
   \addlinespace[2pt]\multicolumn{7}{l}{\textbf{Setting 3}}\\Standalone & -0.00 & 0.00 & -0.00 & -0.00 & -0.00 & $\delta$, $\delta$, 0, 0, 0 \\ 
  Full pooling Bayesian borrowing & -0.00 & -0.00 & 0.17 & 0.17 & 0.16 & $\delta$, $\delta$, 0, 0, 0 \\ 
  Adjacent Bayesian borrowing & -0.00 & -0.00 & 0.11 & -0.00 & -0.00 & $\delta$, $\delta$, 0, 0, 0 \\ 
  RBSB & -0.00 & -0.00 & 0.17 & 0.09 & 0.05 & $\delta$, $\delta$, 0, 0, 0 \\ 
   \addlinespace[2pt]\multicolumn{7}{l}{\textbf{Setting 4}}\\Standalone & -0.00 & 0.00 & -0.00 & -0.00 & -0.00 & $\delta$, $\delta$, $\delta$, 0, 0 \\ 
  Full pooling Bayesian borrowing & -0.00 & -0.00 & -0.00 & 0.18 & 0.18 & $\delta$, $\delta$, $\delta$, 0, 0 \\ 
  Adjacent Bayesian borrowing & -0.00 & -0.00 & -0.00 & 0.11 & -0.00 & $\delta$, $\delta$, $\delta$, 0, 0 \\ 
  RBSB & -0.00 & -0.00 & -0.00 & 0.17 & 0.09 & $\delta$, $\delta$, $\delta$, 0, 0 \\ 
   \addlinespace[2pt]\multicolumn{7}{l}{\textbf{Setting 5}}\\Standalone & -0.00 & 0.00 & -0.00 & -0.00 & -0.00 & $\delta$, $\delta$, $\delta$, $\delta$, 0 \\ 
  Full pooling Bayesian borrowing & -0.00 & -0.00 & -0.00 & -0.00 & 0.18 & $\delta$, $\delta$, $\delta$, $\delta$, 0 \\ 
  Adjacent Bayesian borrowing & -0.00 & -0.00 & -0.00 & -0.00 & 0.11 & $\delta$, $\delta$, $\delta$, $\delta$, 0 \\ 
  RBSB & -0.00 & -0.00 & -0.00 & -0.00 & 0.17 & $\delta$, $\delta$, $\delta$, $\delta$, 0 \\ 
   \addlinespace[2pt]\multicolumn{7}{l}{\textbf{Setting 6}}\\Standalone & -0.00 & 0.00 & -0.00 & -0.00 & -0.00 & $\delta$, $\delta$, $\delta$, $\delta$, $\delta$ \\ 
  Full pooling Bayesian borrowing & -0.00 & -0.00 & -0.00 & -0.00 & -0.00 & $\delta$, $\delta$, $\delta$, $\delta$, $\delta$ \\ 
  Adjacent Bayesian borrowing & -0.00 & -0.00 & -0.00 & -0.00 & -0.00 & $\delta$, $\delta$, $\delta$, $\delta$, $\delta$ \\ 
  RBSB & -0.00 & -0.00 & -0.00 & -0.00 & -0.00 & $\delta$, $\delta$, $\delta$, $\delta$, $\delta$ \\ 
   \addlinespace[2pt]\multicolumn{7}{l}{\textbf{Setting 7}}\\Standalone & -0.00 & 0.00 & -0.00 & -0.00 & 0.00 & $\delta$, $\delta$, $0.6\!\cdot\!\delta$, 0, 0 \\ 
  Full pooling Bayesian borrowing & -0.00 & -0.00 & 0.10 & 0.18 & 0.19 & $\delta$, $\delta$, $0.6\!\cdot\!\delta$, 0, 0 \\ 
  Adjacent Bayesian borrowing & -0.00 & -0.00 & 0.06 & 0.08 & 0.00 & $\delta$, $\delta$, $0.6\!\cdot\!\delta$, 0, 0 \\ 
  RBSB & -0.00 & -0.00 & 0.09 & 0.16 & 0.09 & $\delta$, $\delta$, $0.6\!\cdot\!\delta$, 0, 0 \\ 
   \addlinespace[2pt]\multicolumn{7}{l}{\textbf{Setting 8}}\\Standalone & -0.00 & 0.00 & -0.00 & -0.00 & 0.00 & $\delta$, $\delta$, $0.6\!\cdot\!\delta$, $0.6\!\cdot\!\delta$, 0 \\ 
  Full pooling Bayesian borrowing & -0.00 & -0.00 & 0.10 & 0.09 & 0.20 & $\delta$, $\delta$, $0.6\!\cdot\!\delta$, $0.6\!\cdot\!\delta$, 0 \\ 
  Adjacent Bayesian borrowing & -0.00 & -0.00 & 0.06 & -0.00 & 0.09 & $\delta$, $\delta$, $0.6\!\cdot\!\delta$, $0.6\!\cdot\!\delta$, 0 \\ 
  RBSB & -0.00 & -0.00 & 0.09 & 0.06 & 0.16 & $\delta$, $\delta$, $0.6\!\cdot\!\delta$, $0.6\!\cdot\!\delta$, 0 \\ 
   \addlinespace[2pt]\multicolumn{7}{l}{\textbf{Setting 9}}\\Standalone & -0.00 & 0.00 & -0.00 & -0.00 & 0.00 & $\delta$, $\delta$, $0.6\!\cdot\!\delta$, $0.6\!\cdot\!\delta$, $0.6\!\cdot\!\delta$ \\ 
  Full pooling Bayesian borrowing & -0.00 & -0.00 & 0.10 & 0.09 & 0.08 & $\delta$, $\delta$, $0.6\!\cdot\!\delta$, $0.6\!\cdot\!\delta$, $0.6\!\cdot\!\delta$ \\ 
  Adjacent Bayesian borrowing & -0.00 & -0.00 & 0.06 & -0.00 & -0.00 & $\delta$, $\delta$, $0.6\!\cdot\!\delta$, $0.6\!\cdot\!\delta$, $0.6\!\cdot\!\delta$ \\ 
  RBSB & -0.00 & -0.00 & 0.09 & 0.06 & 0.04 & $\delta$, $\delta$, $0.6\!\cdot\!\delta$, $0.6\!\cdot\!\delta$, $0.6\!\cdot\!\delta$ \\ 
   \bottomrule
\end{tabular}
\caption{Bias for different settings (10\,000 simulations).} 
\label{tab:bias}
\end{table}